\documentclass[12pt,preprint]{aastex}
\usepackage{epsf}

\usepackage{graphicx}
\usepackage{natbib}
\def\old#1{}

\def\oldmath#1{}

\def\abin{a_{\rm bin}}
\def\rclose{r_{\rm close}}

\def\ma{m_1}
\def\mb{m_2}
\def\va{v_1}
\def\vb{v_2}
\def\vej{v_{\rm ej}}

\def\Mbh{M_\bullet}

\def\Msolar{M$_\odot$}
\def\msun{M$_\odot$}

\def\mathMsolar{{\rm M}_\odot}
\def\facR{f_R}
\def\sgr{Sgr~A$^\ast$}
\def\GC{Galactic Center}
\def\kms{km~s$^{-1}$}
\def\AU{{\sc au}}

\begin{document}

\title{Hypervelocity Stars: From the Galactic Center to the Halo}

\author{Scott J. Kenyon}
\affil{Smithsonian Astrophysical Observatory,
\\ 60 Garden St., Cambridge, MA 02138}
\email{skenyon@cfa.harvard.edu}

\author{Benjamin C. Bromley}
\affil{Department of Physics, University of Utah, 
\\ 115 S 1400 E, Rm 201, Salt Lake City, UT 84112}
\email{bromley@physics.utah.edu}

\author{Margaret J. Geller}
\affil{Smithsonian Astrophysical Observatory,
\\ 60 Garden St., Cambridge, MA 02138}
\email{mgeller@cfa.harvard.edu}

\author{Warren R. Brown}
\affil{Smithsonian Astrophysical Observatory,
\\ 60 Garden St., Cambridge, MA 02138}
\email{wbrown@cfa.harvard.edu}

\begin{abstract}

Hypervelocity stars (HVS) traverse the Galaxy from the central black hole to
the outer halo. We show that the Galactic potential within 200 pc acts as
a high pass filter preventing low velocity HVS from reaching the halo. To
trace the orbits of HVS throughout the Galaxy, we construct two forms of the
potential which reasonably represent the observations in the range 5--10$^5$
pc, a simple spherically symmetric model and a bulge-disk-halo model. We use
the Hills mechanism (disruption of binaries by the tidal field of the central
black hole) to inject HVS into the Galaxy and to compute the observable spatial
and velocity distributions of HVS with masses in the range 0.6--4 M$_\odot$.
These distributions reflect the mass function in the \GC, properties of 
binaries in the \GC, and aspects of stellar evolution and the injection 
mechanism. For 0.6--4 \msun\ main sequence stars, the fraction of unbound 
HVS and the asymmetry of the velocity distribution for their bound 
counterparts increases with stellar mass. The density profiles for 
unbound HVS decline with distance from the \GC\ approximately as 
$r^{-2}$ (but are steeper for the most massive stars which evolve off 
the main sequence during their travel time from the \GC); the density 
profiles for the bound ejecta decline with distance approximately as 
$r^{-3}$. In a survey with limiting magnitude $V \lesssim$ 23, the 
detectability of HVS (unbound or bound) increases with stellar mass.

\end{abstract}


\maketitle

\section{Introduction}

The components of the Milky Way sample the diverse forms of matter in 
the universe from the black hole at the dynamical center to the dark 
matter component of the halo. Hypervelocity stars (HVSs), a class of 
recently discovered objects, are a probe linking the central black 
hole and its history to the outer halo. HVSs probably originate from 
interactions with the black hole; they travel out into the halo and 
beyond, serving as tracers of the Galactic potential on scales
from 5 pc to 10$^5$ pc.

To make testable predictions of the relationships between the observable
properties of HVSs and aspects of the Galactic center and halo, we adopt the
massive black hole (MBH)-binary star interaction first proposed by
Hills \citep{hil88} as the injection mechanism for HVSs. We then explore
the impact of the Galactic potential, the properties of binaries in the
Galactic Center, the mass function in the Galactic Center, and stellar
evolution on the observable spatial and velocity distributions of HVSs as
a function of stellar mass.

\subsection {History and Challenges}

\citet {hil88} showed that HVSs are inevitable when 
sufficiently tight binary stars are disrupted by the tidal field of the 
central MBH. Much later, \citet{yu03}  analyzed the production of HVSs 
and argued that the Hills production rates were excessive.  They 
considered other mechanisms for the production of hypervelocity stars
including interactions of single stars with a binary central black hole.

In 2005, \citet{bro05} reported the first discovery of a HVS.  The star,  
SDSS J090745.0+024507, is leaving the Galaxy with a heliocentric radial 
velocity of 853$\pm$12 km~s$^{-1}$ and has the largest velocity ever 
observed in the Milky Way halo. Subsequent photometry revealed that the 
object is a slowly pulsating B main-sequence star \citep{fue06}. Only 
interaction with a massive black hole can plausibly accelerate a 
3 M$_\odot$ main-sequence B star to such an extreme velocity.

The \citet{bro05} discovery inspired further observations along with a 
wealth of theoretical models. \citet{ede05} discovered a hypervelocity 
main sequence B star plausibly ejected from the LMC \citep[see also][]
{bon07} and \citet{hir05} reported a helium-rich subluminous O star 
probably ejected from Sgr A$^*$. These three first discoveries were 
serendipitous.

\citet{bro06a} then carried out a targeted search for hypervelocity stars
by using the SDSS to select faint B-type stars
with lifetimes consistent with travel times from the Galactic center
but that are not a normally expected stellar halo population. So far 
this extensive survey has yielded another 7 hypervelocity stars 
\citep{bro06a, bro06b, bro07a, bro07b}. The velocity of each of these 
stars exceeds the escape velocity from the Milky Way.

The observational discoveries have inspired theorists to reconsider
production mechanisms and ejection rates of hypervelocity stars.
Ejection mechanisms include the original Hills binary encounter with a 
MBH \citep{hil88, yu03, bkg06}, encounters between single stars and 
binary black holes \citep{yu03, ses06, ses07a, ses07b, merritt06}, encounters 
of single stars with an intermediate mass black hole (IMBH) inspiraling 
toward a MBH \citep{han03, gua05, levin06, bau06}, and single star 
encounters with a cluster of stellar mass black holes (SMBHs) around a 
MBH \citep{ole08}. Recently \citet{lu07} predicted hypervelocity ejection 
of tight binary stars in interaction with a central binary MBH. 
Discovery of a single tight binary HVS would be a strong indication 
of a central binary MBH.

Absent the discovery of a binary HVS, distinguishing among the ejection
mechanisms requires predictions of the resulting spatial and velocity
distribution of the HVSs along with computation of expected ejection
rates (e.g. Sesana et al 2006, 2007a). For example, the \citet{hil88} 
mechanism may produce the largest velocities \citep{ses07a} and a binary 
black hole may produce an anisotropic distribution of ejecta \citep{ses06}. 

Dynamical and evolutionary considerations for stellar populations near
the \GC\ also constrain HVS ejection mechanisms. For example, observations
of the current population of B stars at the \GC\ favor the \citet{hil88} 
mechanism over models involving an inspiraling IMBH or a cluster of 
SMBHs \citep{per07b}.

Ejection rate estimates depend on an understanding of the way stars are 
scattered onto orbits intersecting the MBHs ``loss cone'' and on the 
specific ejection mechanism. Perets et al (2007a, 2007b) argue that 
scattering by giant molecular clouds gives HVS production rates from 
binary disruption by a single MBH that are consistent with the observations. 

In addition to the HVSs, all of the ejection mechanisms produce a bound
population of ejecta \citep[e.g.,][]{bkg06}. \citet{bro07a, bro07b} 
demonstrate the very probable presence of a bound population of ejected 
B stars. These stars have velocities between 275 \kms\ and 450 \kms. 
In this velocity range the Brown et al surveys detect 18 outgoing 
stars and only 1 incoming star, consistent with the $\sim$ 200--300~Myr 
lifetime of these main sequence stars and a Galactic center origin 
\citep[see also][]{sve08}.

Because HVSs travel across the Galaxy, they are powerful tracers of the 
Galactic potential \citep{gne05, yu07}. \citet{yu07} emphasize that the
population of bound and unbound ejecta constrain the anisotropy of the 
halo independent of the ejection mechanism.

The power of HVSs as probes of the Galactic center and as tracers of the
Galactic potential provides strong motivation for acquisition of larger
samples of these objects. Brown et al. focus on B stars because these objects
are not an indigenous population in the halo and they are observable to
large distances where the contrast between the density of HVSs and any
indigenous population is largest. 
\citet{koll07} suggest searches for possibly more numerous
old-population, lower mass HVSs. Detection of lower mass
HVSs could provide strong tests of
models for their origin.  

\subsection {From the Galactic Center to the Halo}

In the \citet{hil88} mechanism, HVSs naturally sample both the stellar mass 
function and the properties of the binary population in the Galactic Center. 
The ejecta then traverse the Galaxy and become observable in the halo. We
show that the observable velocity and spatial distributions of HVSs are 
sensitive to the Galactic potential, particularly on scales of 5--200 pc from 
the \GC. These distributions also reflect stellar evolutionary timescales. 

To include as complete a picture of the consequences of the Hills ejection
mechanism as possible, we consider a broad range of
issues in this paper. In \S2.1 we simulate the MBH-binary interaction
following the procedures of  \citet{bkg06}. Motivated by \citet{koll07},
we extend the simulations to cover lower mass stars and unequal mass
binaries. In \S2.2 we derive forms for the Galactic potential which are 
a good representation of the observations from 5--10$^5$ pc. Our results
show that most of the deceleration of HVSs occurs within the central 
200 pc of the Galaxy; thus, a proper fit to the observations in this 
region is a crucial step in understanding HVSs.

In \S2.3, \S2.4, and \S2.5, respectively,  we describe our integration 
of HVS orbits through the Galaxy, the procedure we use to construct mock 
catalogs of HVSs, and the impact of azimuthal deflections on HVS orbits. 

We continue with more detailed discussions of the observable properties 
of HVSs in a simple spherically symmetric potential (\S3) and in a
three-component Galaxy (\S4). One of the most striking general results
of this investigation is that a central potential which matches observations
in the central 200 pc of the Galaxy acts as a high pass filter preventing
lower velocity HVSs from reaching the halo.

In \S3 and \S4 we predict the spatial and velocity distributions of HVSs 
as a function of stellar mass and we address the underlying physical
explanations for the behavior of these distributions. In \S5 we
connect the predictions with the observations by calculating relative space
densities as a function of stellar mass (\S5.1). This discussion
emphasizes that the relative numbers of bound stars are sensitive to the 
mass function in the Galactic Center and relative stellar lifetimes;
relative numbers of unbound stars are nearly independent of stellar lifetime.

In \S5.2 and \S5.3 we analyze some sample HVSs search approaches.  We argue 
that searches become increasingly difficult with decreasing stellar mass 
(\S5.2). We also demonstrate that short lifetimes make the discovery of 
post main sequence HVS unlikely. We conclude in \S6.

\section{The Simulations}

In \citet{bkg06}, we developed initial methods for predicting the observable 
velocity distribution of HVSs in the galactic halo. We derived ejection 
velocities appropriate for a single MBH disrupting binaries with 3--4 \msun\
primary stars and a realistic range of initial binary semimajor axes, $a_{bin}$. 
We then adopted a simple prescription for the Galactic mass distribution, 
integrated the orbits of stars ejected from the \GC, and computed a first 
approximation to the full observable phase-space distribution of ejected 
stars on radial orbits in the halo. This analysis showed that the predicted 
median speeds of ejected stars as a function of distance in the halo are 
consistent with current observations. We also predicted a population of 
ejected stars on bound radial orbits, then discovered by \citet{bro07a} 
as a 3.5$\sigma$ excess of B-type main sequence stars with velocities of 
275--400 km s$^{-1}$.

Here, we extend our analysis to consider a wider range of binary star 
masses and several galactic potentials. Our goals are (i) to understand 
the relative frequencies of HVSs and bound ejected stars as a function of 
stellar mass and (ii) to explore the sensitivity of observable HVS radial
velocities on the Galactic potential from 5 pc to $10^5$ pc.

As in \citet{bkg06}, we simulate the ejection of stars from the \GC\ in 
three steps. Following \citet{hil88,hil92}, we integrate orbits of binary stars 
passing by \sgr\ to quantify ejection probabilities and velocities. Then 
we use a Monte Carlo code based on semianalytical approximations for rapid 
generation of simulated catalogs of ejected stars in the \GC. Finally we 
integrate the orbits of these stars in the Galactic potential to calculate 
how they populate the Galaxy's halo.

To generate observable samples, the simulation incorporates the stellar main 
sequence lifetime from published stellar evolution calculations.  We depart
from \citet{bkg06} by considering primary and secondary stars with masses 
$\ma$ and $\mb$ in the range 0.6--4~\msun. In addition to considering orbits
of stars in a simple spherical Galaxy, we also derive velocity distributions for 
stars ejected into a standard Galactic potential consisting of a bulge, disk, 
and extended halo.

\subsection{Massive Black Hole--Binary Star Interaction} 

To obtain the spectrum of ejection speeds as a function of the
properties of the disrupted binary star, we first simulate the 
MBH-binary interaction. As in \citet{bkg06}, we use a sixth-order, 
symplectic integrator \citep{yos90, broken06, kenbro06} to simulate 
the disruption of a binary system falling into an MBH with mass 
$\Mbh = 3.5 \times 10^6$~\msun\ \citep[e.g.][]{ghe05}. The 
binaries have a random orbital axis and phase; the semimajor 
axis is in the range $a_{bin}$ = $(a_{min},~a_{max})$, where $a_{min}$ 
depends on the binary mass and $a_{max} \approx$ 4 AU. We assume that 
the center of mass of the binary is on a hyperbolic orbit with an 
initial approach speed of 250 km s$^{-1}$ \citep{hil88} and that the 
binary has negligible orbital eccentricity.

Motivated by \citet{koll07}, we extend our simulations to equal mass 
binaries with lower mass primary stars and to unequal mass binaries 
(Table 1). Table 1 includes the main sequence lifetime $t_{ms}$, 
the stellar radius at the time of core hydrogen exhaustion $R_{ms}$, 
and the minimum $a_{bin}$ for a binary with unit mass ratio, 
$q \equiv \ma / \mb$ = 1 \citep{schaetal92,schaetal93}. 
To minimize the probability of a collision during the encounter 
($\lesssim$ 10\%), we set $a_{min}$ = 2 $R_{ms,1} / r_L$ 
\citep[see also Table 1 of][]{gin07}, where $R_{ms,1}$ is 
the radius of the primary star and
\begin{equation} \label{eq:rl}
r_L = \frac{0.49 q^{2/3}}{0.6 q^{2/3}~+~ln~(1 + q^{1/3})}
\end{equation}
is the fractional radius of the inner Lagrangian surface for a binary 
system with a circular orbit \citep{egg83}. This assumption allows us to 
sample a wide range of binary systems without worry that the interaction 
with the central black hole leads to tidal interaction between the binary 
components and possible coalescence \citep{gin07}. Table 1 also lists 
$a_{min}$ for unequal mass binaries with 0.6 \msun\ secondary stars.

Our calculations reproduce published results and they are consistent with 
analytical estimates of HVS production \citep{hil88,hil92,yu03,gouqui03}. 
The median ejection velocity -- measured at infinity -- is
\begin{equation}\label{eq:vej}
\vej = 1760 
\left(\frac{\abin}{\rm 0.1\ AU}\right)^{-1/2} 
       \left(\frac{\ma + \mb}{2\ \mathMsolar}\right)^{1/3}
\left(\frac{\Mbh}{3.5 \times 10^6\ \mathMsolar}\right)^{1/6}
       \facR \ \   {\rm km~s^{-1}} ~ , ~
\end{equation}
where $m_1$ is the mass of the primary star and 
$m_2$ is the mass of the secondary star. The factor $\facR$ 
is a normalization factor that depends on 
$r_{close}$, the distance of closest approach to the black hole:
\begin{eqnarray}\label{eq:facR}
\nonumber
    \facR & = & 0.774+(0.0204+(-6.23\times 10^{-4}
	   +(7.62\times 10^{-6}+ \\
& &
(-4.24\times 10^{-8}
	+8.62\times 10^{-11}D)D)D)D)D,
\end{eqnarray}
where
\begin{equation}\label{eq:D}
D =
\left(\frac{\rclose}{\abin}\right)
	\left[\frac{2 \Mbh}{10^6 (\ma + \mb)}\right]^{-1/3}.
\end{equation}
This factor also sets the probability for an ejection, $P_{ej}$:
\begin{equation}
\label{eq:PE}
P_{ej} \approx 1-D/175
\end{equation}
for $0 \le D \le 175$. For $D > 175$, $\rclose \gg \abin$; the binary 
does not get close enough to the black hole for an ejection and 
$P_{ej} \equiv 0$.

For any binary system, the ejection probability and median ejection
speed are set by the binary parameters -- $a_{bin}$, $\ma$, and $\mb$ --
and the distance of closest approach to the black hole $r_{close}$.
We assume that we can ignore the chance of a merger event between 
binary companions and that ejections of the primary and secondary 
star are equally likely.  Because they can only occur in rare cases where 
$\abin$ and $\rclose$ are both small \citep[e.g., also][]{gin06,gin07},
mergers are unlikely to change our results by more than 10\%.

Our simulations suggest that the binary mass ratio affects the ejection 
probability only when the mass ratio is extreme ($\ma/\mb > 10$). There 
is then a consistent preference for ejection of the secondary 
\citep[for additional details, see][]{bkg06}. When the 
binary contains unequal mass stars, the median ejection speeds of the 
primary and secondary stars are 
\begin{equation}\label{eq:vavb}
\va = \vej \left(\frac{2\mb}{\ma+\mb}\right)^{1/2}  \ \ \ {\rm and} \ \ \
\vb = \vej \left(\frac{2\ma}{\ma+\mb}\right)^{1/2} ,
\end{equation}
respectively.  In all cases, the distribution of ejection speeds is 
approximately Gaussian with a dispersion of 20\% of the median speed.  

\subsection{The Galactic Potential}

Once ejected stars leave the central black hole, they travel out through the 
Galaxy and decelerate. Historically, mass models for the Galaxy developed 
for other applications are optimized to fit observations at distances 
$r \gtrsim$ 200 pc from the \GC\ \citep[e.g.,][] {deh98,klyp02,yu03,wid05}.  
For ejected stars, however, the largest deceleration occurs in the central 
200 pc (see Figure \ref{fig:gaxmodacc} below). Thus, we need a Galaxy model 
that reproduces the Galactic potential within 200 pc of the \GC.

In their analysis of the proper motion of Arches, a cluster of young stars 
in the \GC, \citet{sto07} also realized that standard Galactic potential 
models are inadequate to address dynamical phenomena in the \GC. They 
derive a three component triaxial model for the central Galaxy which 
allows them to investigate the orbits of Arches cluster stars in the 
central potential.  However, their application does not depend on an 
accurate potential beyond the \GC; thus, their parameters do not match 
the potential at $r \gtrsim$ 0.5--1 kpc. 

Because HVSs traverse the Galaxy, we need a potential that fits observations 
from the \GC\ to the outer halo.  Thus, we derive Galaxy models that fit both
the observations at 5--200 pc (where HVSs decelerate considerably) and at 
10--100 kpc (where we observe HVSs today).

\subsubsection{Parameterization of the Galactic potential}

We consider two parameterizations of the Galaxy potential. First, we follow 
\cite{bkg06} and adopt a simple, spherically symmetric density profile: 
\begin{equation}\label{eq:rho}
\rho(r) = \frac{C}{1 + (r/r_c)^2} ~ .
\end{equation}
To yield a mass of $\sim 3\times 10^7$~\msun\ within 10 pc of the \GC\
\citep[as inferred from stellar kinematics;][]{eck97,lau02,sch03,ghe05}
and a circular rotation speed of 220~km~s$^{-1}$ in the Solar neighborhood
\citep{hog05}, we adopt\footnote{Although \citet{bkg06} 
used the value for $C$ quoted here in their numerical simulations, they 
quoted $C = 1.27 \times 10^4$ in the text. Here, we quote the correct
value for $C$.}
$r_c = 8$~pc and $C = 1.396 \times 10^4$~\Msolar pc$^{-3}$.
This density profile yields a mass of $\sim 1.2 \times 10^9$ \msun\ inside 
120 pc, compared to an estimate of $8 \pm 2 \times 10^8$ \msun\ in the 
nuclear stellar disk \citep{lau02}.  The corresponding potential is 
\begin{equation}
\Phi_G(r) =  - 2\pi G C r_c^2 ~ [2~(r_c/r)~\arctan(r/r_c)~+~\ln(1~+~r^2/r_c^2)], 
\label{eq:phi1}
\end{equation}
where $G$ is the Gravitational constant. If we add the potential of the
central black hole, the total potential is
\begin{equation}
\Phi(r) = \Phi_G(r) - G \Mbh / r 
\end{equation}

To examine the impact of $\Phi$ on our results, we also consider 
a three component, spherically symmetric potential that includes 
contributions from the bulge ($\Phi_b$), disk ($\Phi_d$), and halo ($\Phi_h$) 
\citep[e.g.,][]{gne05,wid05}. 
Specifically, we adopt
\begin{equation}\label{eq:phi}
\Phi_G = \Phi_b + \Phi_d + \Phi_h ~ ,
\end{equation}
where
\begin{eqnarray}\label{eq:phis}
\nonumber
\Phi_b(r) & = & - G M_b / (r + a_b), \\
\Phi_d(R,z) & = & - G M_d /\sqrt{R^2 + [a_d + (z^2+b_d^2)^{1/2}]^2}, \ \ \rm{and} \\
\nonumber
\Phi_h(r) & = & - G M_h \ln(1+r/r_s) / r
\end{eqnarray}
\citep[e.g.,][]{her90,miy75,nav97}.
Here, $r$ is the radial coordinate in a spherical coordinate system
and $(R,z)$ are cylindrical coordinates.  

To choose appropriate parameters for this potential, we consider 
multi-component mass models for the Galaxy.  Most derivations adopt analytic 
functions for the mass density, $\rho(R,z)$, and then derive fitting parameters 
from least-squares fits to a set of dynamical observables \citep[e.g.,][]
{deh98,klyp02,bat05,deh06}.  \citet{wid05} expand on this technique with an 
iterative approach that solves for a stable, self-consistent bulge-disk-halo 
model which fits the standard observables \citep[see also][]{wid08}.  
Although both of these approaches fit the data well for \GC\ distances 
ranging from $\sim$ 200 pc to $\sim$ 100--200 kpc, they are not optimized 
for the \GC\ region at $\sim$ 5--200 pc \citep[see also][]{sto07}. 

Here, we derive a potential model that provides a reasonable match to
observations throughout the range 5--$10^5$ pc.  Table 2 lists our 
results. We follow \citet{gne05} and adopt disk\footnote{To maintain a
circular velocity of 220 km~s$^{-1}$ at 8 kpc, we reduce the disk scale
length from 5 kpc to 4 kpc.} and halo parameters from \citet{klyp02}. 
For the bulge, we derive parameters that fit velocity dispersion data 
at 5--200 pc \citep{tre02}, the mass within 10 pc inferred from stellar 
kinematics \citep{eck97,lau02,gen03,sch03,ghe05}, and the mass of the 
nuclear stellar disk inside 120 pc 
\citep[$8 \pm 2 \times 10^8$ \msun;][]{lau02}. Using a Newton-Raphson 
method that over-weights measurements in the central 100 pc, we derive 
$M_b = 3.55 \times 10^9$ \msun\ and $r_b$ = 106 pc.

This model for the bulge underestimates the stellar mass within a few 
pc of the \GC\ \citep{gen03}. Because mass in the central 10--100 pc 
has the largest impact on HVS trajectories, this underestimate has 
little impact on our results. 

Table 2 compares the parameters for our Galaxy model with other models
used for calculations of HVS trajectories.  \citet{yu07} adopt the 
\citet{klyp02} parameters for the bulge and disk and the \citet{die07} 
halo model. For the \citet{deh98} multi-component triaxial potential,
we derived the best-fitting spherical, three component model for the 
radial component of their potential. Our mass and scale length for the
bulge agree well with the \citet{deh98} model. 

Table 3 compares the mass-radius relations for several Milky Way models
with observations. For the observed masses -- listed as `Target' in the
Table -- enclosed within 10 pc ($M_{10}$) and 120 pc ($M_{120}$), we adopt 
the kinematic mass from \citet{ghe05} and the mass of the nuclear stellar 
disk from \citet{lau02}. 
For the mass enclosed within $r$ = 210 pc ($M_{210}$), we use the measured 
velocity dispersion at 210 pc \citep[136 \kms;][]{tre02,wid05} and make the 
conversion $M_{210} = 3 r v^2 / G$. At larger radii, we adopt the circular 
rotation speed of 220 \kms\ at $r$ = 8 kpc \citep[for $M_{8000}$;][]{hog05}
and the \citet{klyp02} Milky Way mass at 100 kpc (for $M_{100000}$). Aside
from a low value at 210 pc, our mass model provides a good match to the 
target values. The \citet{sto07} results provide a better match at 210 pc,
but underestimate the mass at 10 pc. Most other mass models \citep[e.g.,][]
{deh98,gne05,yu07} underestimate the mass substantially for all $r \lesssim$ 
210 pc. The \citet{wid05} models match the target data outside $\sim$ 120 pc, 
but they overestimate the mass inside 100 pc (see their Fig. 6).

\subsubsection{Impact of the central potential on HVS trajectories}

\begin{figure}[htb]
\centerline{\includegraphics[width=7.0in]{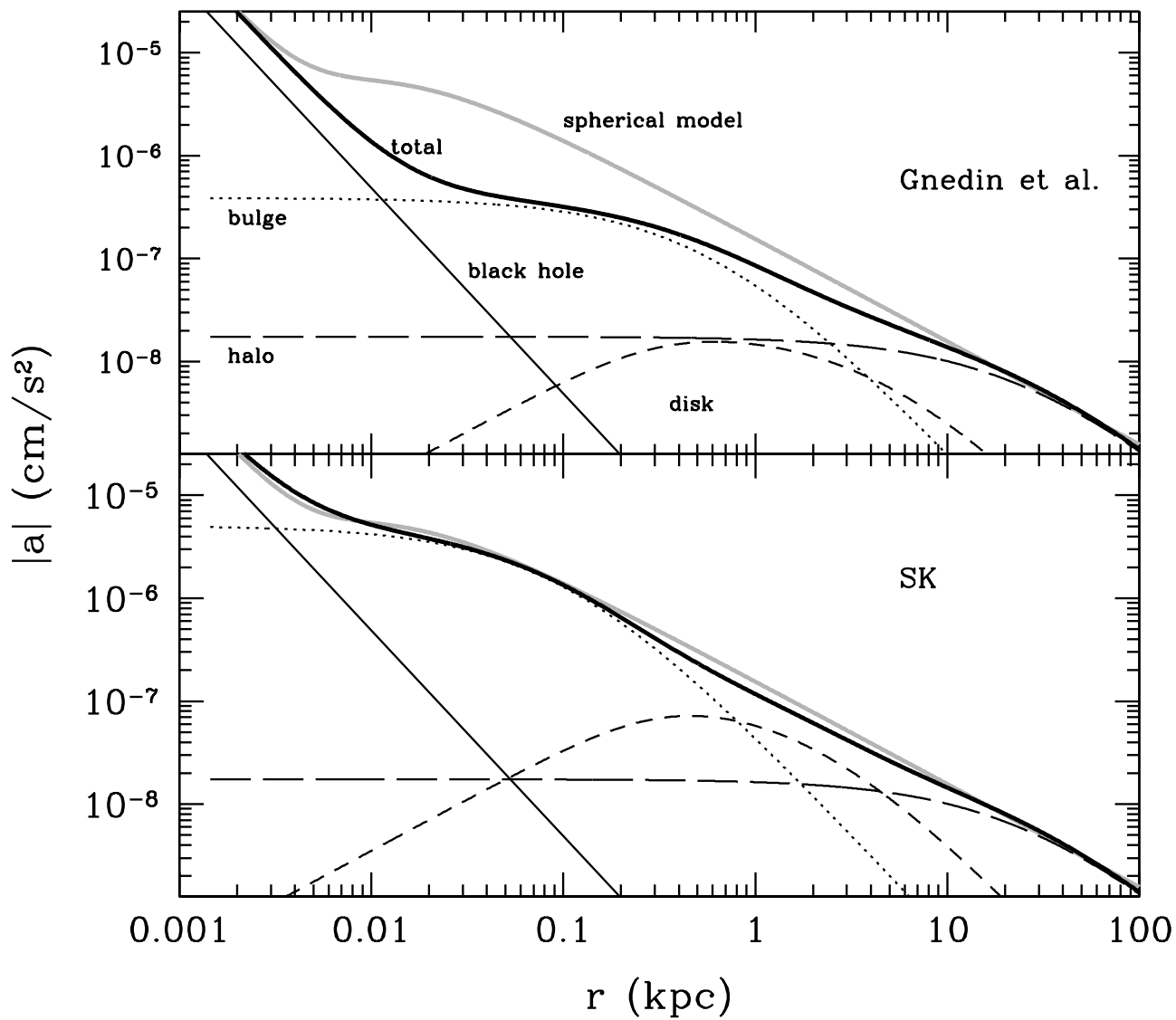}}
\vskip -5ex
\caption{Acceleration in the Galactic potential. 
Lower panel: decomposition of our Galaxy potential into contributions 
from the central black hole (thin solid black line), bulge (dotted line), 
disk (short dashed line), and halo (long dashed line). The thick solid
line is the combined potential for comparison with the simple spherical
potential (thick grey line).
Upper panel: as in the lower panel but for the \citet{gne05} Galaxy
potential.
\label{fig:gaxmodacc}}
\end{figure}

The radial variation of the acceleration, $a = \Phi / r$, demonstrates 
the importance of the central 200 pc of the Galactic potential in the 
ejection of HVSs (Figure \ref{fig:gaxmodacc}). For $r \lesssim$ 4 pc, 
the central black hole dominates the deceleration of ejected stars 
\citep{tre02,wid05}.  At larger radii, the acceleration has a clear 
plateau with a radial width comparable to the scale length $r_b$ and 
then falls rapidly. For $r \gtrsim$ 5--10 $r_b$, the disk and halo 
provide most of the deceleration.  Although the disk adds to the halo 
component at $r \sim$ 1--20 kpc, its contribution is always relatively 
small.

To illustrate how the acceleration varies among different potentials,
Figure \ref{fig:gaxmodacc} compares $|a|$ for our simple spherical 
potential with the three component model adopted here and a three 
component model developed to analyze HVS trajectories in a triaxial 
Milky Way halo \citep{gne05}. Most other Galaxy potentials have
acceleration profiles similar to the \citet{gne05} potential 
\citep[e.g.][]{deh98,klyp02,bat05,deh06,yu07}. Because the 
\citet{wid05} potential overestimates the mass in the \GC, this
model overestimates $|a|$ in the \GC.

The upper panel of Figure \ref{fig:gaxmodacc} compares $|a|$ for the 
\citet{gne05} potential with our simple spherical model. The larger 
$r_b$ for the \citet{gne05} potential yields a plateau with smaller 
acceleration ($|a| \sim 5 \times 10^{-7}$ cm s$^{-2}$) at $r \sim$ 
20--200 pc. Other At $r \gtrsim$ 1--100 kpc, the acceleration from 
the disk and halo match the simple potential.

The lower panel of Figure \ref{fig:gaxmodacc} compares $|a|$ from our 
simple spherical model to $|a|$ for our adopted three component Galaxy. 
The relatively small scale length of our bulge yields a good match to 
the plateau ($|a| \sim 6 \times 10^{-6}$ cm s$^{-2}$) of the spherical 
model at small radii, $r \sim$ 5--30 pc. The disk + halo potential matches 
the spherical model well at large distances, $r \gtrsim$ 1--2 kpc. 

\begin{figure}[htb]
\centerline{\includegraphics[width=7.0in]{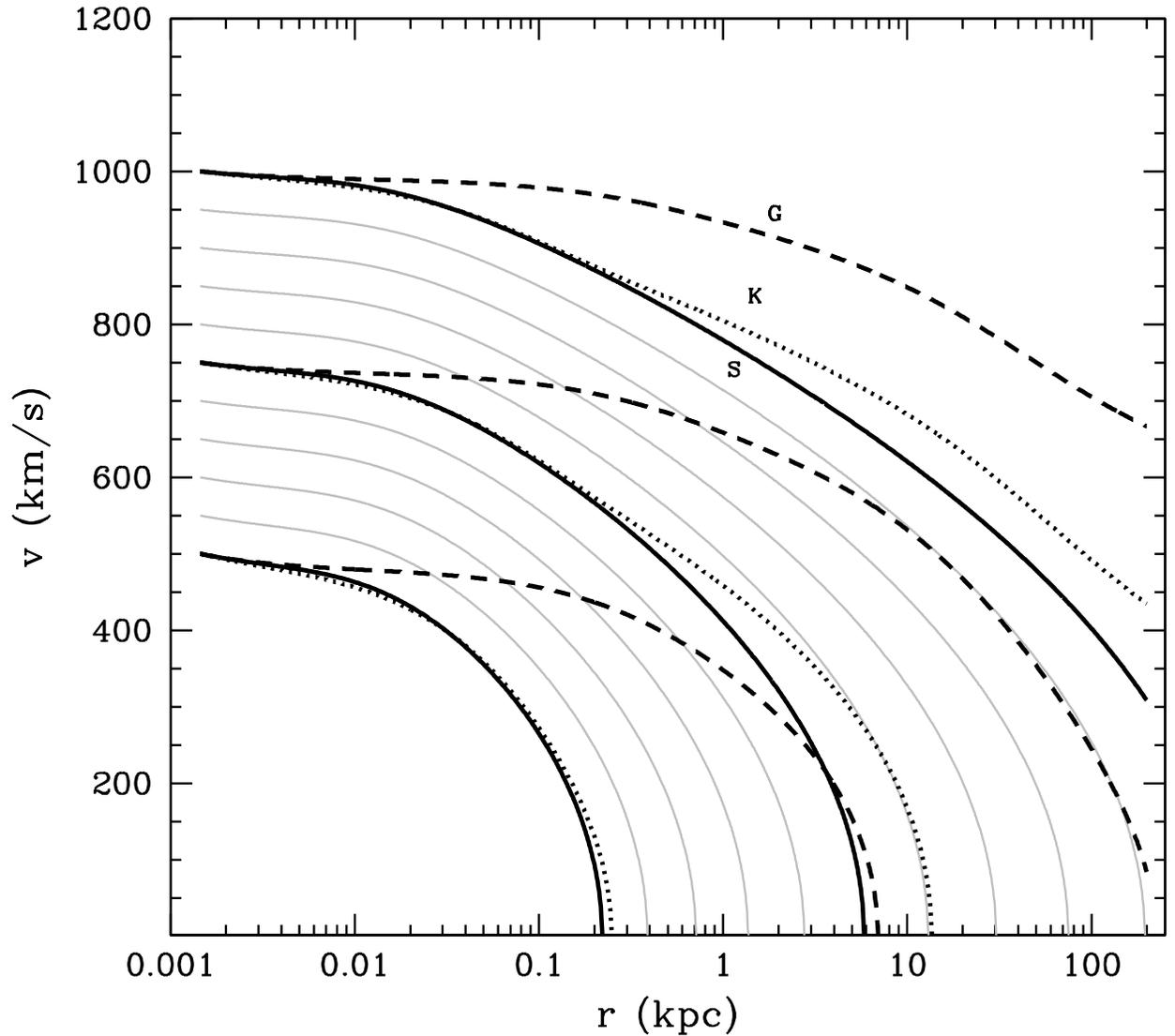}}
\caption{Radial velocity profiles of ejected stars in different Galaxy 
potentials. The labels indicate profiles in the \citet{gne05} potential (G), 
our three component potential (K), and the simple spherical potential (S).
The differences in the three sets of profiles starting at $v_0$ = 500, 
750, and 1000 \kms\ illustrate the impact of the central potential in
setting the observable properties of HVS in the Galactic halo.
\label{fig:gaxmodvel}}
\end{figure}

The acceleration profile of the inner Galaxy has a clear impact on
the ejection of HVSs into the outer Galaxy (Figure \ref{fig:gaxmodvel}).
The solid lines in Figure \ref{fig:gaxmodvel} show $v(r)$, the variation 
of velocity with Galactocentric distance for the simple spherical potential 
as a function of $v_0$, the initial ejection velocity at 1.45 pc.
The figure also shows several results for our three component Galaxy
(dotted curves) and for the \citet{gne05} potential (dashed curves). 
For the simple spherical potential, ejected stars with $v_0 \lesssim$ 
625 \kms\ fail even to reach $r \sim$ 1 kpc. Stars with 625 \kms\ 
$\lesssim v_0 \lesssim$ 800 \kms\ reach $r \sim$ 1--10 kpc;
stars with $v_0 \gtrsim$ 800 \kms\ reach $r \gtrsim$ 10 kpc. We derive
similar results for our adopted three component Galaxy. In the \citet{gne05} 
potential, stars with relatively small ejection velocities, 525 \kms\ 
$\lesssim v_0 \lesssim$ 750 \kms\ can reach large halo distances, 
$r \sim$ 10--100 kpc. Thus, the larger deceleration produced by a 
more realistic potential for the \GC\ leads to smaller $v(r)$ and 
less penetration of ejected stars into the halo.

From these results, we expect clear differences between the observable
velocity histograms of stars ejected into the \citet{gne05} potential 
and the histograms of stars ejected into our three component Galaxy model.  
We discuss this issue in more detail in \S3--4.

\subsection{Integration of Orbits through the Galaxy}

To generate populations of stars ejected from the \GC, we perform Monte 
Carlo simulations of $\sim 10^6$ stars on radial orbits starting a small 
distance away from the black hole \sgr. Following \citet{bkg06}, we 
assume $a_{bin}$ is distributed with equal likelihood per logarithmic
interval \citep[see also][]{abt83,duq91,hea98} between $a_{min}$ (Table 1) 
and $a_{max} \approx$ 4~AU. 
Consistent with the expectation of strong gravitational focusing of
the orbit of the center of mass of the binary by \sgr, we choose the 
distance of closest approach, $\rclose$, from a distribution that varies 
as $\rclose$. For an adopted pair of stellar masses $\ma$ and $\mb$,
the random $a_{bin}$ and $\rclose$ yield the ejection probability,
$P_{ej}$. If $P_{ej}$ exceeds a random deviate, we choose an ejection 
velocity $v_0$ from a gaussian distribution with average velocity 
$v_{ej}$ and dispersion 0.2 $v_{ej}$.

To follow the trajectories of these ejected stars through the Galaxy, 
we use a simple leap-frog integrator \citep{bkg06}. We start each star
on a radial trajectory with velocity $v_0$ at a distance of 1.45 pc 
from \sgr. For our adopted potential, the mass in stars within this 
radius from \sgr\ is roughly 5\% of the black hole's mass 
(Table \ref{tbl:MWMass}). Thus, our results are fairly independent of 
the starting position. Once the initial trajectory is set, we use the 
leap-frog integrator to derive the time-dependent velocity of the star 
through a Galaxy with potential $\Phi$ (Eq. \ref{eq:phi1} or Eq. \ref{eq:phi}). 

Simulations of HVSs with several potentials provide a better understanding 
of the physical processes involved in ejecting stars from the \GC\ into 
the Galactic halo.
Calculations with the simple spherical potential (Eq. \ref{eq:phi1}) allow 
us to isolate the impact of stellar evolution from the deceleration of 
ejected stars traveling through the Galaxy (\S3). Simulations with the
three component Galaxy (Eq. \ref{eq:phi}) yield clear estimates for the 
sensitivity of observable quantities on the mass in the \GC\ and the 
triaxiality of the potential (\S4).

\subsection{Predicting Observable Quantities}

To generate mock catalogs of HVSs, we assume the central black hole 
ejects stars at a constant rate on timescales comparable to the typical
orbital timescales of the bound population, $t_{orb} \sim 1$ Gyr. 
We eject each star at a time $t_{ej}$, randomly chosen between zero and 
the main sequence lifetime of the primary star $t_{ms,1}$, and observe 
each star at a time $t_{obs}$, randomly chosen between zero and the main 
sequence lifetime for the ejected star ($t_{ms,ej}$). The first constraint 
($t_{ej} < t_{ms,1}$) guarantees that the binary interacts with the 
central black hole before post-main sequence evolution disrupts the 
binary. The second constraint ($t_{obs} < t_{ms,ej}$) guarantees that 
Earthbound observers detect the ejected star before it evolves off the 
main sequence.  If $t_{ej} < t_{obs}$, we assign an ejection speed $v_0$ 
to the ejected star and integrate its orbit through the Galaxy for a 
travel time $t_t = t_{obs} - t_{ej}$. For each ejected star with index
$i$ observed at time $t_{obs,i}$, the mock catalog includes the derived 
distance $r_i$, the initial ejection velocity $v_{0i}$, the radial velocity 
at time $t_{obs}$ $v_i$, and the orbital elements for bound stars.

Although this procedure for generating catalogs of HVSs differs from 
the method described in \citet{bkg06}, it yields similar results. 
Motivated by the high star formation rate in young clusters near the \GC\ 
\citep[e.g.,][]{fig99,fig02,lang05}, we previously adopted a short timescale 
between formation and ejection e.g., $t_{ej} \ll t_{ms,1}$ \citep{bkg06}.
Because binary stars outside nearby star forming regions can also encounter 
the black hole \citep[e.g.,][]{per07}, here we adopt a less restrictive 
constraint on the ejection timescale ($t_{ej} < t_{ms,1}$). Direct 
comparisons of new catalogs with the \citet{bkg06} catalogs show negligible 
differences in the predicted median speeds as a function of distance from
the \GC\ or in the velocity histograms of halo stars (see below). Thus, our 
simulations are relatively insensitive to assumptions concerning the 
ejection time from the Galactic Center\footnote{We assume a constant Galactic
potential.  Because the Galaxy probably gains mass with time 
\citep[e.g.,][]{bul05}, our estimates for the bound population of 
older, less massive stars are more uncertain than estimates for 
more massive stars.}

\subsection{Azimuthal Deflection of HVS Trajectories}

With main sequence lifetimes longer than a typical orbital timescale 
$t_{orb} \sim$ 1 Gyr, bound stars with masses smaller than 1.5 \msun\ can 
make several passes through the \GC\ and, perhaps, interact with the 
central black hole or other stars near the \GC\ before returning to 
the halo. For the spherically symmetric potential of Eq.~\ref{eq:rho}, 
we estimate that bound stars on trajectories that pass within 10~pc 
of \sgr\ have less than a 0.01\% chance of passing within 10~\AU\ of 
another star in the Galactic Center. The probability of encountering 
a star outside of the 10~pc region is even smaller. With speeds 
$\sim$ 800--1000~km~s$^{-1}$ near the \GC, a bound ejected star must 
pass within 0.1 AU of another star to suffer a significant deflection 
of its orbit. Thus, bound ejected stars do not change their orbits in 
a spherically symmetric Galaxy \citep[see also][]{yu07}.

For the three component Galactic potential, significant deflections of HVS 
trajectories occur only if the potential is not spherical.  Although a 
triaxial halo potential can produce significant proper motions in a sample of 
HVSs \citep{gne05}, these produce little change in the radial component
of the velocity in the halo.  After several passages through the Galactic 
Center,  however, \citet{yu07} note that a triaxial halo may redistribute 
a bound star's kinetic energy between the radial and angular velocity. 
For reasonable triaxial halo potentials, the observed radial velocities 
of ejected bound stars change by $\sim$ 10\% \citep{yu07}.

\begin{figure}[htb]
\centerline{\includegraphics[width=7.0in]{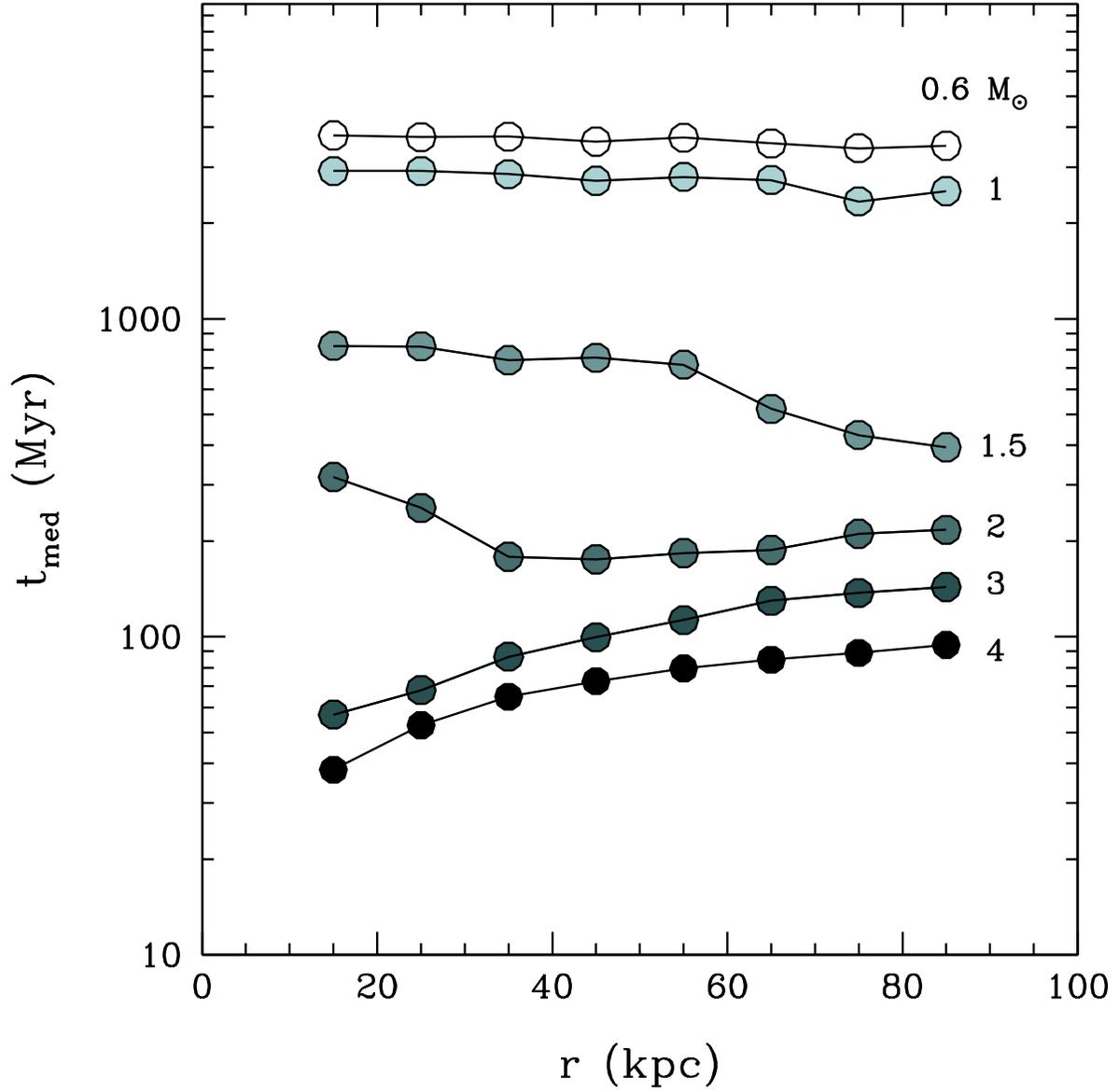}}
\caption{Median age of ejected stars on their first pass through
the halo as a function of Galactocentric distance. The labels indicate 
the masses of the ejected stars.
\label{fig:medage}}
\end{figure}

A triaxial bulge potential can have a more dramatic impact on the observed 
velocities of ejected and bound HVSs. Current data suggest the mass distribution
near the \GC\ is not spherical \citep[see][and references therein]{deh98,klyp02,
deh06,sto07}. Because it is easier to eject stars along the minor axis of a
triaxial \GC\ potential (e.g., Fig \ref{fig:gaxmodacc}--\ref{fig:medage}), 
we expect 
potentially large differences in the observable velocities of HVSs as a 
function of galactic latitude. To estimate the magnitude of this effect, we 
derive predicted velocity distributions for ejected and bound stars in two 
different three component Galaxy potentials below (\S4).

\section{Observable Ejected Stars in a Simple Spherically Symmetric Potential}

To isolate how the important physical effects of the ejection process
relate to observable properties of ejected stars, we begin with a discussion 
of the evolution of ejected stars in a simple, spherically symmetric Galaxy 
(Eq. \ref{eq:rho}). We describe the evolution in the three component Galactic 
potential in \S4. We further restrict our analysis of observable properties 
to stars with $r \ge$ 10 kpc, where current surveys can distinguish ejected 
stars from the indigenous halo population and where the observed radial 
velocity is nearly indistinguishable from the true radial velocity of 
ejected stars \citep[e.g.,][]{bro06a,bro06b,bro07a,bro07b}.

We divide these successful ejected stars into bound stars that orbit 
the \GC\ and unbound stars that leave the Galaxy. Because our forms 
for the Galactic potential do not yield true escape velocities, we
define unbound stars as ejected stars that reach $r \gtrsim$ 200 kpc
with $v \gtrsim +200$ \kms\ (see Figure \ref{fig:gaxmodvel}). These stars 
have $v \gtrsim +450$ \kms\ at $r =$ 50 kpc and $v \gtrsim +300$ \kms\ 
at $r =$ 100 kpc.

With typical orbital 
periods of 1 Gyr, bound stars falling back into the \GC\ must have 
long lifetimes. Thus, bound stars are either low mass stars or massive 
stars on their first pass out through the Galactic halo. For unbound 
stars, the typical travel times to $r \sim$ 40--80 kpc are 
$\sim$ 100 Myr \citep{bro07b}. Thus, some 3--4 \msun\ unbound stars 
with lifetimes of 150--350 Myr evolve off the main sequence while
traveling through the halo. This evolution removes potential HVSs from
the catalogs, reducing the relative frequency of massive stars at 
$r \gtrsim$ 50--100 kpc (see below).  Stellar evolution removes very
few low mass stars from the catalogs; thus, the observable properties 
of unbound 0.6--2 \msun\ stars depend solely on the binary orbital 
parameters and the Galactic potential.

Figure \ref{fig:medage} shows the median ages for stars ejected from 
equal mass binaries on their first pass outward through the halo. 
Stars with radial speeds exceeding the escape velocity dominate the 
population of massive stars at all distances from the \GC. Thus, the 
observed median ages of 3--4 \msun\ stars increase monotonically with $r$. 
Bound stars dominate the population of low mass stars at all 
Galactocentric distances. Thus, the median ages of 0.6--1 \msun\ stars 
are roughly constant with $r$. For intermediate mass stars, the median 
ages represent a mix of mostly bound stars at small $r$ and mostly 
unbound stars at large $r$.  The distribution of median ages for 
2 \msun\ stars is dominated by bound stars for $r \lesssim r_u 
\approx$ 20 kpc and by unbound stars for $r \gtrsim r_u$, where
$r_u$ is the Galactocentric distance where the fraction of unbound
stars exceeds 10\%--15\%. For 1.5 \msun\ stars, the bound population 
dominates for $r \lesssim r_u \approx$ 70--80 kpc. Thus, the median 
ages for 1.5--2 \msun\ stars fall for $r \lesssim r_u$ and rise or 
remain constant for $r \gtrsim r_u$.

At large Galactocentric distances, the radial distribution of ejected stars 
depends on the stellar mass and the shape of the Galactic potential
(Fig. \ref{fig:densallhex}). Unbound stars (dashed lines in the Figure)
are ejected at the largest possible velocities.  They are the stars least 
affected by the Galactic potential and thus have shallow density profiles.
For long-lived, low mass stars, the density profile of unbound stars is 
slightly steeper than the $\rho \propto r^{-2}$ expected for a point-like 
potential.  Bound stars are ejected at smaller velocities and reach
smaller Galactocentric distances than unbound stars. Thus, bound stars 
have steeper density profiles.  For both bound and unbound stars, low mass 
stars with main sequence lifetimes longer than the travel time from the \GC\ 
have shallow density profiles.  Although they are typically ejected at 
larger velocities, many shorter-lived 3-4 \msun\ stars evolve off the main 
sequence before reaching 50--100 kpc and thus have the steepest density 
profile, particularly at large $r$. 

The large range in main sequence lifetimes among 0.6--4 \msun\ stars leads
to clear differences in the fraction of unbound stars as a function of
stellar mass and Galactocentric distance (Fig. \ref{fig:fracunbound}). 
Massive main sequence stars observed at $r \gtrsim$ 50 kpc must have relatively 
large ejection speeds and short travel times. Thus most massive stars at 
large $r$ are unbound.  Low mass main sequence stars ejected on bound 
trajectories have much longer to reach $r \gtrsim$ 50 kpc before evolving 
off the main sequence. Because there are more ejected stars on bound orbits
than on unbound orbits, most observable ejected low mass stars are bound 
at all $r$.

The velocity distributions of observable ejected stars are also a strong 
function of
stellar mass (Fig. \ref{fig:fountainhex}; Table 2). For 3--4 \msun\ stars, 
the velocity histograms are asymmetric with median velocities of $\sim$
400--500 km s$^{-1}$ for stars with $r \gtrsim$ 30 kpc. The median
velocity is sensitive to $r$: nearer stars have smaller median
velocities. Because the typical travel time to $r \gtrsim$ 30 kpc is 
a significant fraction of $t_{ms}$ for a 3--4 \msun\ star, more massive
stars must have larger ejection speeds to reach larger $r$. Thus, more 
distant regions of the halo tend to hold faster-moving ejected stars
\citep{bkg06}.

For stars with masses $\lesssim$ 2 \msun, the velocity histograms are
more symmetric with decreasing stellar mass. Two physical effects produce 
symmetric velocity histograms for the lowest mass ejected stars: (i) most 
ejected stars that reach $r \gtrsim$ 10 kpc are bound and (ii) because 
1 \msun\ stars live far longer than 2--4 \msun\ stars, most of the 
bound population consists of low mass stars (Fig. \ref{fig:fracunbound}). 
For typical $t_{orb} \sim$ 1 Gyr, these stars live long enough to make 
several orbits of the Galaxy. Bound 2--4 \msun\ stars nearly always 
evolve off the main sequence before completing a single orbit of the 
Galaxy. Thus, bound stars increasingly dominate the velocity histograms 
of observable lower mass stars.

To illustrate the importance of the stellar lifetime in more detail, 
Figs. \ref{fig:vvsr}--\ref{fig:voutvsr} show the median speed as a 
function of Galactocentric distance for the entire observable population 
(Fig. \ref{fig:vvsr}) and for outgoing observable stars only 
(Fig. \ref{fig:voutvsr}). The 3--4 \msun\ stars ejected at relatively 
small velocities ($v_0 \lesssim$ 700--800 \kms; Fig. \ref{fig:gaxmodvel}) 
do not live long enough to reach $r \gtrsim$ 10 kpc. Lower mass stars live 
long enough for nearly all ejected stars to reach $r \gtrsim$ 10 kpc. Thus,
more massive stars observed at $r \gtrsim$ 10 kpc move faster than lower 
mass stars. At even larger distances, the short stellar lifetimes of massive 
stars select for stars ejected at even larger velocities. Thus, the median 
speed of observable massive ejected stars is a strong function of $r$ 
\citep[see also][]{bkg06}. The longer lifetimes of low mass stars allows a 
larger fraction of ejected stars to reach large $r$; thus, the median speed 
depends only weakly on $r$.

The median speeds of primary stars ejected from unequal mass binaries are
fairly insensitive to the mass ratio $q$ of the disrupted binary 
(Fig. \ref{fig:uneqmvvsr1}). Massive primary stars ejected from low mass 
binaries tend to have somewhat smaller median speeds than massive stars 
ejected from more massive binaries.  For binaries with 0.6 \msun\ 
secondaries, the large mass ratio leads to the smallest ejection velocities 
and thus the longest travel times for the 4 \msun\ primary 
(Eq. \ref{eq:vavb}). For binaries with 1 \msun\ secondaries, the 
longer travel times lead to smaller median speeds only for $r \gtrsim$ 80 kpc, 
where the density profile is very steep (Fig. \ref{fig:densallhex}).

The median speeds of secondary stars are much more sensitive to $q$ (Fig. 
\ref{fig:uneqmvvsr2}). Although secondary stars ejected from unequal mass 
binaries are ejected at larger speeds than their equal mass counterparts
(Eq. \ref{eq:vavb}), two features of the ejection process lead to 
{\it smaller} median speeds for secondary stars observed at 
$r \sim$ 10--80 kpc (compare with Fig. \ref{fig:vvsr}).  
Because the primary stars of unequal mass binaries have shorter lifetimes 
than the secondaries, the median age of an ejected secondary star is 
smaller for an unequal mass binary than for an equal mass binary. Thus, the 
secondary stars in unequal mass binaries have longer median travel times.
Longer travel times allow secondaries ejected at velocities larger than the 
median ejection velocity (Eq. \ref{eq:vej}) to reach $r \gtrsim$ 80 kpc, 
removing these stars from the population at $r \sim$ 20--80 kpc. Longer 
travel times also allow more slowly moving ejected stars to reach large $r$. 
Removing the highest velocity stars from the population and adding more 
slowly moving stars to the population reduces the median velocities of 
secondaries at $r \sim$ 20--80 kpc.

To summarize, simulations in the simple spherical potential yield two
classes of ejected stars observable at $r \gtrsim$ 20 kpc. The bound 
population of mostly long-lived, low mass stars has a steep radial 
density profile, $\rho(r) \propto r^{-n}$ with $n \approx$ 3, and a 
symmetric radial velocity profile centered at $v_r \approx$ 0 \kms
(Figs. \ref{fig:densallhex}--\ref{fig:fountainhex}; Table \ref{tbl:vmeds}).  
Short-lived 3--4 \msun\ bound stars have slightly steeper radial density 
profiles and provide a high-velocity tail to the radial velocity distribution 
at +275 to +500 \kms.  Unbound ejected stars with $v \gtrsim$ +500 \kms\
have shallow radial density profiles ($n \approx$ 2--2.5). The short 
lifetimes of 3--4 \msun\ unbound stars yield very steep density profiles 
at large radii ($n \gtrsim$ 3 for $r \gtrsim$ 80 kpc). Thus, the relative
frequency of 3--4 \msun\ unbound stars decreases relative to lower mass
unbound stars at $r \gtrsim$ 80 kpc.

\section{Observable Ejected Stars in a Three-Component Galaxy} 

We now consider the evolution of observable ejected stars in a three 
component potential consisting of a bulge, disk, and halo (Eq. \ref{eq:phi}). 
We discuss results for the \citet{gne05} Galaxy model, which has a 
relatively small acceleration at small $r$, and for our three component
model Galaxy, which has a larger acceleration at small $r$.  The acceleration 
at small $r$ for our model Galaxy is comparable to the acceleration for 
the simple spherical potential (\S2.2; Fig. \ref{fig:gaxmodacc}).  To 
minimize differences in median velocities caused by the relatively massive 
disk in our model Galaxy, we derive results for ejections along the z axis.
To minimize the number of low velocity stars ejected into the halo of the
\citet{gne05} model Galaxy, we derive results for ejections in the disk 
plane, where the larger deceleration from the disk prevents more low
velocity ejected stars from reaching 10 kpc.

At 10--60 kpc, the velocity histograms derived for ejections into our 
model Galaxy
and the simple spherical Galaxy are indistinguishable (Figure \ref{fig:compmod}).
In both cases, the distributions for 4 \msun\ stars are greatly skewed to large
velocities, with more low velocity stars at 10--30 kpc than at 30--60 kpc. These
models yield a small ($\sim$ 10\%) fraction of stars returning to the \GC\ with
negative velocities.  For 0.6 \msun\ stars, the distributions are nearly symmetric 
about zero velocity, consistent with a large population of bound stars in both 
examples.  In addition to a modest fraction ($\sim$ 10\%) of unbound stars, 
models for 0.6 \msun\ stars have a large fraction ($\sim$ 45\%) of returning stars.

At 10--60 kpc, calculations with the \citet{gne05} Galaxy potential yield more symmetric 
velocity histograms and many fewer unbound stars than calculations using the other 
Galaxy models.  For 2--4 \msun\ stars, the velocity histograms are skewed to high 
velocities, but have a smaller fraction of stars with velocities exceeding
600--800 \kms\ (Figure \ref{fig:compmod}). Ejections into this potential also produce
a much larger population of stars at 10--30 kpc than at 30--60 kpc.  For lower mass 
stars, the velocity histograms are more similar to those derived from the other Galaxy
models. However, velocity histograms with this potential have smaller dispersions
about the median velocity.

Table \ref{tbl:vmeds} compares median speeds at 30--60 kpc for the three model
Galaxies. For 2--4 \msun\ primary and secondary stars, the median speeds derived
for our three component Galaxy are $\sim$ 5\% to 15\% smaller than those derived
for the simple spherical Galaxy. This difference results from the potential at
$r \sim$ 1 kpc, where the three component potential produces a smaller acceleration
than the simple potential (Fig. \ref{fig:gaxmodacc}). The median speeds of massive
stars in the \citet{gne05} potential are typically $\sim$ 75\% of those for the other
potentials. For 0.6--1 \msun\ stars, the median speeds derived for the three models 
are similar.  Although the median velocities of low mass stars in the \citet{gne05} 
potential are $\sim$ 2/3 the median velocities derived for the other potentials, the 
typical difference of 6--12 \kms\ is small compared to current observational errors.

In these calculations, the large differences between the median speeds and the 
shapes of the velocity histograms are set by the Milky Way potential at $r$ = 
5--200 pc (Figure \ref{fig:gaxmodacc}). At small $r$, the potential acts as a 
high pass filter, preventing low velocity stars from reaching the halo (Figure
\ref{fig:gaxmodvel}).  In the \citet{gne05} potential, lower velocity stars
reach the halo. Thus, the median speeds of observable stars are smaller and 
the velocity histograms
are narrower.  In the simple spherical Galaxy and our three component Galaxy,
low velocity stars are trapped at $r \lesssim$ 10 kpc. Thus, the median speeds
are larger at all $r \gtrsim$ 10--20 kpc. Because low velocity bound stars
remain at $r \lesssim$ 10 kpc in our three component Galaxy, the velocity 
histograms of these stars are also broader for all $r \gtrsim$ 10--20 kpc.

Because the potential of the central 200 pc of the Galaxy makes such a 
large difference in the median speeds of observable HVSs at 10--60 kpc, 
our results suggest that HVSs might be a useful probe of the triaxial 
potential of the central Galaxy.  To test this possibility, we consider 
two triaxial models for the inner Galaxy. 
\citet{sto07} analyze the mass density data of \citet{lau02} and derive a three 
component triaxial model for the central 200 pc. Each component in this model 
has axial ratios $x_0$ : $y_0$ : $z_0$. With a minimum axial ratio of 0.71 in 
the z-direction (for the nuclear stellar disk), we predict an expected reduction 
of 15--20 \kms\ in the median speeds of stars ejected along the z-axis relative 
to stars ejected in the plane of the Galaxy.  Using OGLE data for the bulge, 
\citet{rat07} derive a triaxial model for the Galactic bar with $x_0:y_0:z_0$ = 
1.0 : 0.35 : 0.26. Assuming that the rotation of the bar eliminates any signal in 
HVS speeds along the y-axis, we expect a reduction of 25--30 \kms\ in the median 
speeds of HVSs ejected along the z-axis of the bar relative to those ejected in 
the Galactic plane.

Given current uncertainties in mass models for the Galaxy, our simulations 
suggest that the radial density profiles and velocity histograms of HVSs at 
$r \gtrsim$ 20 kpc are most sensitive to the form of the Galactic potential 
for $r \lesssim$ 0.1--1 kpc. At $r \gtrsim$ 1 kpc, small changes to the 
potential produce negligible differences in observable quantities. At 
$r \lesssim$ 200 pc, triaxial potentials consistent with observations
produce $\pm$ 10--30 \kms\ differences in the median velocities of 
observable unbound 
stars as a function of galactic latitude and negligible differences in the 
radial density profiles.  These differences are comparable to the variation
in median speeds among 0.6--1.5 \msun\ ejected stars, where the shorter
stellar lifetimes of more massive stars produce median speeds offset from
$v_r \approx$ 0. With high quality stellar mass estimates, HVSs could 
provide constraints on the triaxiality of the central galactic potential. 

\section{Scaled Density Profiles and Searches for HVS}

The \citet{hil88} ejection mechanism for HVSs has several observable
consequences. In \citet{bkg06} and in this paper, we show that radial 
velocity surveys for massive main sequence stars should reveal 
comparable numbers of bound and unbound HVSs.  Our results also 
suggest that 4 \msun\ HVSs should have smaller space densities 
at $r \sim$ 50--100 kpc than 3 \msun\ HVS. Observations confirm 
both predictions.  \citet{bro07a,bro07b} identify 26 bound B-type 
HVSs with $v_r \gtrsim$ 275 \kms\ and only 1 bound HVS with 
$v_r \lesssim -275$ \kms, supporting the notion that the B-type HVSs 
are main sequence stars and are not horizontal branch stars.  In these 
data, longer-lived 3 \msun\ HVSs fill the survey volume out to $r \sim$ 
80--90 kpc; shorter-lived 4 \msun\ HVSs are missing at large distances.
This result confirms our expectation that stellar lifetimes set the
HVS space density at large $r$.

From an analytic analysis of the radial density profiles of unbound stars, 
\citet{koll07} develop a strong motivation to search for HVSs among 0.6--1 
\msun\ stars. They conclude that a survey of faint stars near the main 
sequence turnoff might yield as many as 0.022 HVSs deg$^{-2}$, roughly a 
factor of 20 larger than the $\sim$ 0.001 HVSs deg$^{-2}$ inferred from 
our targeted survey of B-type stars \citep{bro07b}. Our calculations in 
\S3--4 suggest that the large bound population of observable ejected turnoff
stars overwhelms the unbound component, especially at $r \sim$ 10--40 kpc 
where the radial velocities of turnoff stars can be measured with existing
multi-objects spectrographs (Fig. \ref{fig:fountainhex}).  However, if the 
relative space density of bound, low mass ejected stars is large enough
relative to the indigenous halo population, fruitful searches for low
mass HVSs might still be possible \citep[see][]{koll07}.
The success of the HVS ejection model for 3--4 \msun\ stars and the
\citet{koll07} suggestion leads us to take a more detailed look at the 
likelihood of identifying large samples of HVSs among lower mass stars.

\subsection{Relative Space Densities as a Function of Stellar Mass}

To quantify the relative space densities of HVSs as a function of stellar
mass in our simulations, we must scale the density profiles in Fig. 
\ref{fig:densallhex}.  For each stellar mass, we require the relative
frequency of primary and secondary stars as a function of stellar mass 
(the mass function), the relative frequency of binaries with the range 
of separations $(a_{min}, a_{max})$, and an efficiency factor relating
the number of stars ejected from the \GC\ to the number observed at 
$r \approx$ 10--100 kpc.  For simplicity, we assume that the binary 
frequency is independent of primary mass and that the mass function is 
independent of orbital separation. Although observations suggest the 
binary frequency depends on stellar mass \citep[e.g.,][]{lad06}, the 
differences in binary frequencies among nearby 0.6--4 \msun\ stars are 
small compared to uncertainties in the mass function and the relative 
frequency of binary separations. 

To construct a first estimate of the scaled density profiles of unbound 
stars as a function of stellar mass, we develop a simple model. For each 
stellar mass $m$, we derive a weighting factor 
\begin{equation}
x(m) = x_1(m) ~ x_2(m,a) ~ x_3(m) ~ t_{ms}
\label{eq:xxx}
\end{equation}
where $x_1(m)$ is the relative number of stars of mass $m$,
$x_2(m,a)$ is the relative number of binary stars with orbital 
separations in the range $a_{bin} = (a_{min}, a_{max}$), and 
$x_3(m)$ is the relative fraction of stars ejected from 1.45 pc 
that reach $r \gtrsim$ 10 kpc.  To account for stellar evolution, 
we scale the weighting factor by the main sequence lifetime
$t_{ms}$ in Table \ref{tbl:MSBin}.  We set $x(m$ = 0.6 \msun) = 1 
and normalize other weights accordingly. This model assumes that 
primary and secondary stars are selected from the same mass function, 
a reasonable first approximation.

To derive $x_1$, we adopt a simple power law mass function
\begin{equation}
\xi(m) dm \propto  m^{-(q+1)} dm
\label{eq:imf}
\end{equation}
with $q \approx$ 1--1.5 \citep[e.g.][]{sal55,ms79}. Integrating this 
function over $m$ yields relative numbers of stars in mass ranges 
with lower mass limits $m_l$ and upper mass limits $m_u$. For a set 
of stellar masses, $S(m) = \{0.6, 1.0, 1.5, 2.0, 3.0, 4.0 \}$, we 
choose a set of lower mass limits 
$S(m_l) = \{0.4, 0.8, 1.25, 1.75, 2.5, 3.5 \}$ and
a set of upper mass limits
$S(m_u) = \{0.8, 1.25, 1.75, 2.5, 3.5, 4.5 \}$.

To guide our choices for the exponent $q$, we rely on recent observations 
of the Arches cluster near the \GC\ \citep{sto05,kim06}. These analyses 
suggest a current mass function with $q \approx 0.9$, top-heavy
compared to the local Salpeter mass function with $n \approx$ 1.35. 
However, dynamical interactions among cluster stars may produce mass 
segregation on timescales comparable to the cluster age, flattening 
the mass function in the core and steepening the mass function in the 
outskirts of the cluster \citep{dib07,port07}. Thus, the most likely
exponent for Arches stars is probably $q \approx$ 1.0--1.35.  To provide 
estimates for a range of $q$, we derive 
$S(x_1) = \{1., 0.29, 0.13, 0.086, 0.05, 0.025 \}$ for $q$ = 1.35 and
$S(x_1) = \{1., 0.36, 0.18, 0.14, 0.09, 0.05 \}$ for $q$ = 1.0.

Observations of the Arches cluster and other young stars currently 
provide no constraints on the frequency of separations for close 
binaries in the \GC. To derive $x_2$, we thus rely on observations
in the local solar neighborhood.  For nearby binaries with A-type and 
B-type primary stars, $a_{bin}$ is distributed with roughly equal 
likelihood per logarithmic interval \citep[][]{abt83,hea98}. For stars 
with orbital separations in the range $a_{bin} = (a_{min}, a_{max}$) 
from Table 1, this assumption yields a set of values for $x_2$, 
$S_1(x_2) = \{1., 0.66, 0.45, 0.41, 0.39, 0.39 \}$.  

For nearby solar-type stars, the frequency of binary separations is
closer to a log-normal distribution \citep{duq91}. For this relation,
we derive $S_2(x_2) = \{1., 0.95, 0.90, 0.87, 0.86, 0.85 \}$. 
Because this frequency distribution under-weights binaries with small 
separations, it under-weights low mass ejected stars relative to massive
ejected stars. To maximize the predicted relative density of low mass 
stars, we adopt the $S_1$ set of weights for $x_2$.  This decision may
overestimate the relative abundance of solar-type stars\footnote{Other 
physical processes -- including binary evaporation near the MBH 
\citep{per07b} and interactions between binaries and molecular clouds 
\citep{per07} -- can lower the relative abundance of solar-type stars 
among HVS. To maximize the predicted density of low mass stars among 
HVS, we ignore these processes in our estimate of $x_2.$} by a 
factor of $\sim$ 2

To derive $x_3$, we rely on the calculations for \S3--4. For a complete
derivation of $x_3$, we require a full set of simulations spanning the
entire range of binary separations and mass functions for primary and
secondary stars. Because very few binaries produce ejected stars, this
simulation requires a significant computational effort for little return. 
In the set of simulations made for this paper, the ratio of failed to 
successful ejections is nearly independent of the masses of the primary 
and secondary stars and the initial binary separation. Thus, for our 
first estimate of $x$, we adopt $x_3$ = 1. 

Fig. \ref{fig:densrelquad} shows the scaled relative density profiles 
of observable main sequence stars from this simple model.  For clarity, 
we normalize the density profiles for 1--4 \msun\ stars to the density
profiles for 0.6 \msun\ stars. Thus, the relative densities of 0.6 \msun\ 
stars are unity at all $r$.  For all 0.6--4 \msun\ ejected stars (upper 
panels), the density profiles scale inversely with stellar mass. Thus, 
low mass ejected stars are relatively more abundant than massive stars 
at all $r$. Because shallower mass functions contain relatively more 
massive stars, the differences in the scaled density profiles are smaller 
for smaller $n$.  For unbound 1--4 \msun\ stars at $r \sim$ 10--200 kpc, 
the density profiles are roughly constant with $r$ and scale inversely 
with stellar mass (lower profiles). Aside from the dramatic decline in
the relative density of 4 \msun\ stars at $r \gtrsim$ 30 kpc due to their
relatively short main sequence lifetimes, massive unbound stars are a 
factor of $\sim$ 10 less abundant than low mass unbound stars at all $r$.

Two physical effects cause the changes in the slopes of the scaled 
density profiles in Fig. \ref{fig:densrelquad}. Because massive ejected
stars contain a larger fraction of unbound stars (Fig. \ref{fig:densallhex}),
massive ejected stars have shallower density profiles at small $r$ than
low mass ejected stars. Thus, the scaled density profiles for massive
stars rise with $r$. At large $r$, however, short stellar lifetimes
remove massive stars from the sample of observable unbound stars. For
4 \msun\ stars, the typical travel time to $r \sim$ 50--100 kpc is
comparable to the main sequence lifetime, $t_{ms} \approx$ 160 Myr
(Table 1). Thus, the relative densities of 4 \msun\ stars decline
considerably at $r \sim$ 50--100 kpc. For 2--3 \msun\ stars, the
typical travel time to $r \sim$ 50--100 kpc is less than half of the
main sequence lifetime, $t_{ms} \approx$ 300 Myr to 1 Gyr. Thus, the
relative densities for these stars decline with $r$ less dramatically
than the relative densities for massive stars.

The relative numbers of bound stars as a function of stellar mass depend 
primarily on the mass function and the relative stellar lifetimes (Eq. 
(\ref{eq:xxx})). For a mass function with $q$ = 1.35 and the stellar
lifetimes in Table \ref{tbl:MSBin}, the relative number of 0.6 \msun\ 
and 3 \msun\ stars is $\sim$ 800:1, close to the prediction of $\sim$
1000:1 from the detailed model. 

In contrast, the relative numbers of unbound stars are nearly independent
of stellar lifetime. Because the time for unbound stars to reach $r \sim$ 
50--100 kpc is a small fraction of the main sequence lifetimes for 0.6--2
\msun\ stars, the relative numbers of 0.6--2 \msun\ unbound stars depends
primarily on the adopted mass function. Thus for $q$ = 1.35, the relative 
number of 0.6 \msun\ and 2 \msun\ unbound stars is $\sim$ 12:1. For more 
massive stars, the travel time is comparable to the main sequence lifetime. 
If we assume that $\sim$ 50\% of 3--4 \msun\ stars evolve off the main 
sequence as they travel from 10 kpc to 100 kpc, the relative numbers of 
0.6 \msun\ and 3--4 \msun\ unbound stars is $\sim$ 50:1 for a mass function
with $q$ = 1.35, close to the prediction in Fig. \ref{fig:densrelquad}. 

\subsection{Predicted Detections}

Converting the relative density profiles of Fig. \ref{fig:densrelquad}
into predicted detection rates requires a targeting strategy. We follow 
\citet{bro06a} and assume target stars are selected from a magnitude-limited 
survey. For simplicity, we consider a shallow magnitude-limited survey 
to $V_{max}$ = 21 and a deeper survey to $V_{max}$ = 23. To estimate 
relative detection rates, we derive absolute $V$ magnitudes as a 
function of stellar mass using the \citet{schaetal92} and 
\citet{schaetal93} evolutionary tracks and bolometric corrections 
from \citet{kh95}. We use these estimates to derive the maximum 
observable distance, $r_{max}(m,V_{max})$, for each stellar mass 
in our set of masses.  Integrating the relative densities in 
Fig. \ref{fig:densrelquad} over distance from $r$ = 8 kpc to 
$r_{max}(m,V_{max})$ then yields the relative numbers of stars of
each mass as a function of $V_{max}$. For surveys that yield one 
observable 4 \msun\ HVS, we predict $n_{e,m}$ the relative number 
of observable ejected stars at mass $m$ and $n_{u,m}$ the number 
of observable unbound stars at mass $m$.  Table \ref{tbl:RelNumHVS} 
lists our results.

Shallow surveys with $V_{max} \le$ 21 are sensitive to HVSs of all
masses (see columns (3) and (5) of Table \ref{tbl:RelNumHVS}).  
Roughly 40\% (12\%) of the unbound (bound) ejected stars are 
2--4 \msun\ stars with $M_V$ = $+1.7$ to $-1$.  Because lower mass 
stars have longer main sequence lifetimes, these stars comprise a 
larger fraction of bound stars in a magnitude-limited survey. 
Nearly all ($\sim$ 94\%) observable ejected stars with $m \approx$ 
0.6--1.5 \msun\ are bound to the Galaxy. However, $\sim$ 25\% of 
ejected 2--4 \msun\ stars are unbound.  Thus, shallow surveys 
targeting 2--4 \msun\ stars have a higher probability of 
observing unbound HVS than surveys for lower mass stars.

Deeper surveys with $V_{max}$ = 23 yield larger fractions of lower
mass stars in the bound and unbound populations (see columns (4) and 
(6) of Table \ref{tbl:RelNumHVS}). Although our analysis predicts 
that $\sim$ 50\% (75\%) of the unbound (bound) ejected stars with 
$V \le$ 23 are 0.6--1 \msun\ stars, only $\sim$ 5\% of observable
low mass ejected stars are unbound. Among 1.5--4 \msun\ stars, we 
derive a much larger fraction of unbound stars, $\sim$ 25\%. Because
3--4 \msun\ unbound stars are relatively rare in surveys with $V_{max}$
= 23, deep surveys targeting 1.5--2 \msun\ stars have the highest 
probability of detecting unbound HVS.

Several competing physical effects combine to produce the results in 
Table \ref{tbl:RelNumHVS}. Massive stars are much brighter than low
mass stars; thus, magnitude-limited surveys probe much larger volumes
for massive stars than for low mass stars. For typical density profiles 
$\rho \propto r^{-2}$, the relative numbers of stars scale with their
limiting distances $r_{max}$ \citep[see also][]{koll07}.  All unbound 
stars spend comparable times traveling from 10 kpc to 200 kpc. Thus, 
the relative numbers of unbound stars scale with the mass function at
the \GC, $r_{max}$, and the ratio of travel time to stellar lifetime.  
For 1 \msun\ stars and 3 \msun\ stars, these factors yield relative 
numbers of 6:1 (mass function), 1:10 ($r_{max}$), and 2:1 (stellar
lifetimes); the simple estimate for the relative number of $\sim$ 1:1 
is close to the tabulated values. 

To make a simple estimate for the relative numbers of bound stars, we 
need a larger correction for relative stellar lifetimes.  Because low 
mass stars live longer than the typical orbital time of $\sim$ 1 Gyr, 
the bound population samples low mass stars ejected throughout the lifetime 
of the Galaxy.  For 1 \msun\ and 3 \msun\ stars, the relative stellar 
lifetimes ($\sim$ 30:1) yield relative numbers of 15:1 in a magnitude
limited survey, close to the tabulated values of 10--20:1. Thus, as 
\citet{koll07} first pointed out, low mass bound stars dominate the 
population of observable ejected stars in all magnitude limited surveys.

As a final piece of this analysis, we consider identifying HVSs within 
the large population of indigenous halo stars. \citep{bro06a} use SDSS 
$(g^\prime -r^\prime)_0$ and $(u^\prime - g^\prime)_0$ colors to target
3--4 \msun\ HVS with $g^\prime_0 \lesssim$ 19.5.  Searching for massive 
stars in this color-magnitude space has several advantages over searches 
for lower mass HVSs: (1) surveys for bright, massive HVSs sample a large 
volume, (2) the predicted fraction of unbound stars is $\sim$ 50\%
(Fig. \ref{fig:densallhex}), (3) the predicted velocity histogram for 
bound stars is more asymmetric (Fig. \ref{fig:fountainhex}) and thus they
are more readily detectable, and (4) two colors reduce white dwarf and 
quasar contamination considerably, yielding a 3 \msun\ HVS candidate 
list of manageable size \citep[see also][]{bro07a,bro07b}. 

In a 6800 square degree survey of 575 candidates with 
$17.0 < {g^\prime}_0 < 19.5$, \citet{bro07b} identify 7 unbound HVSs and 
9 probable bound HVSs. The surface number density of unbound 3 M$_\odot$
HVS is $\sim$ 0.001 per square degree; the surface number density of 
observed 3 M$_\odot$ ejected stars is $\sim$ 0.002 per square degree.
In this shallow survey, the relative numbers of bound and unbound HVS
agree well with model predictions \citep{bro07b}.  The survey detection 
efficiency (number of HVS detected / number of candidates) is 1.2\% for 
unbound objects.

At the other end of the mass range we explore in Table \ref{tbl:RelNumHVS},
\citet{koll07} propose observing fainter ($19.5 < g^\prime < 21.5$) stars 
near the main sequence turnoff with masses $m \approx$ 0.8--1.3 \msun.
Applying scaling arguments to the \citet{bro06a} detections, \citet{koll07}
predict 1 turnoff HVS per 45 square degrees within their suggested survey 
limits. These arguments, and a later corroborating estimate by \citep{bro07b},
assume that the velocity distribution of HVSs is independent of their mass, 
contrary to the more detailed calculations for Table \ref{tbl:RelNumHVS}.

However, \citet{koll07} scale the relative numbers of massive HVSs by 
the present day mass function, which is a product of the initial mass 
function (IMF) and stellar evolution\footnote{Formally, we derive a 
crude estimate of the present day relative mass function as 
$x_1(m) ~ t_{ms}$ in Eq. (\ref{eq:xxx}).}. This approach is valid for 
the bound population.  Thus, the predictions of 1 turnoff HVS per 
45--50 square degrees \citep{koll07,bro07b} are reasonable estimates 
for the bound population of low mass HVSs.  Because all unbound stars 
spend roughly equal amounts of time traveling through the halo, the 
relative numbers of unbound stars scale with the IMF. This scaling 
reduces the predicted sky surface density of unbound low mass HVSs 
relative to unbound massive HVSs as outlined above. Thus, \citet{koll07} 
overestimate the detectable population of unbound HVSs by a factor of 
$\sim$ 30--100. Unbound stars are a small fraction of the observable 
ejected turnoff stars.

According to Table \ref{tbl:RelNumHVS}, $\sim$ 5\% of turnoff stars in 
a survey for low mass HVS with $g_0^{\prime} \le$ 21.5 are unbound. 
Within the more dominant bound population, $\sim$ 20\% should have 
$|v| \gtrsim$ 275 \kms. For a predicted sky surface density of 1 bound
HVS per 45--50 square degrees, detecting unbound or bound low mass HVS 
within a larger (300 per square degree) population of indigenous halo 
stars may be challenging. Detecting a single unbound low mass HVS
requires observations of $\sim 3 \times 10^5$ stars. This large sample 
would yield $\sim$ 5 bound candidates with $|v| \gtrsim$ 275 \kms. 
We emphasize that only the signature of an unbound HVS is unambiguous; the 
bound component may be difficult to distinguish from the well-populated
tails of the velocity distribution of the indigenous population.

These estimates are sensitive to the ejection model for HVS and
to the adopted mass function and the fraction of stars in binaries 
with separations ($a_{min}, a_{max}$) in our model ($x_1$ and $x_2$ 
in Eq. (\ref{eq:xxx})). As outlined above, both of these inputs to 
our model are uncertain. 

Nonetheless, these models indicate that
identification of HVS is increasingly difficult as the mass of the star
decreases. The drivers of this conclusion are that (1) in a magnitude limited
survey massive stars are visible to greater depth, (2) the fraction of unbound
stars increases with mass, (3) the velocity distributions are increasingly
symmetric for less massive HVS, and (4) the indigenous contamination increases
with decreasing stellar mass.

If HVS are ejected by a binary MBH instead of a single MBH, we expect 
different scaling laws with stellar mass \citep[e.g.,][]{ses06,ses07a}. 
Clearly, for the \citet{hil88} mechanism, the properties of binaries 
in the Galactic center are important; they are irrelevant for ejection 
by a binary MBH. Thus, searches for low mass HVS provide tests of 
basic inputs to our estimates for the \citet{hil88} ejections mechanism 
as well as the overall picture for the origin of HVS \citep[e.g.,][]{per07b}. 

\subsection{Post Main Sequence HVS}

Although our results indicate that low mass main-sequence stars are not
prime targets for dedicated HVS surveys, we must consider whether their
post-main sequence descendants -- horizontal branch (HB) and red giant 
branch (RGB) stars -- are more favorable targets.  HB stars are as luminous
as 2--4 \msun\ main sequence stars and are thus observable to larger
distances than their 1--2 \msun\ progenitors \citep[see][]{bro07b}. 
With larger luminosities than HB stars, RGB stars are also reasonable 
targets for HVS surveys.

To quantify the probabilities for detecting HB or RGB HVSs, we derive
$r_{max}(m,V)$ from adopted absolute magnitudes and lifetimes. For HB
stars, we adopt $M_V \approx$ 0.5 \citep{bro08} and lifetimes $\sim$
1\% of the main sequence lifetime \citep{yi01}. Scaling the relative
numbers by the relative lifetimes (a factor of 0.01) and the relative
volumes (a factor of 7--8), the HB descendants of 1--2 \msun\ main 
sequence stars are a factor of $\sim$ 10 less abundant than their main
sequence progenitors. Thus, HB stars are unlikely to provide many HVSs
in a deep survey.

For RGB stars, we consider deep optical and near-infrared (IR) surveys.
We adopt an absolute brightness at the tip of the RGB 
\citep[$M_V \approx -1.5$; $M_K \approx -6$;][]{bell01} and a typical 
lifetime of 10 Myr \citep{yi01}. For optical surveys with limiting
magnitudes $V \le $ 21--23, we predict that RGB stars are $\sim$ 2\%
as abundant as 1--2 \msun\ main sequence stars. In an IR survey with 
a depth of K = 15 \citep[e.g., 2MASS][]{skr06}, we predict that RGB 
stars are $\lesssim$ 1\% as abundant as 1--2 \msun\ main sequence 
stars in a deep optical survey. Although RGB stars are observable 
to distances of 100--200 kpc in deep IR and optical surveys, their 
short lifetimes preclude detection as HVSs. 

\section{Conclusions}

HVSs are a fascinating newly discovered class of objects because they 
connect the Galactic Center with the outer halo of the Milky Way. We 
explore these connections by using the \citet{hil88} model to inject 
stars into the Galactic potential. We track the journeys of these
ejected stars across the Galaxy and derive simulated catalogs of 
observable HVSs. 

The foundation for our model includes the construction of forms for the
Galactic potential which fit observations over the range 5--10$^5$ pc.
We demonstrate that potentials which match the observations within 
the central 200 pc of the Galaxy are crucial for understanding HVSs. 
Our approximations to the potential may be useful for other astrophysical 
problems which connect the central regions of the Galaxy to its outer reaches.

We show that important aspects of the median speeds and shapes of the
observable velocity distributions of HVSs are set by the Milky Way potential
at r $\lesssim$ 200 pc. For the potentials we construct to match the 
observations in this central region, median speeds are larger and 
velocity histograms are broader. Thus, low velocity ejected stars have 
less penetration into the outer halo at every stellar mass. These results 
indicate that HVSs might be useful probes of the triaxial potential 
of the central Galaxy.     

The models predict the spatial and velocity distributions of observable HVSs.
They also provide a physical understanding of the origin of the dependences 
of these distributions on stellar mass for $m =$ 0.6--4 \msun\ and on 
distance in the halo for $r =$ 10--200 kpc. Here, we enumerate the main 
predictions of the model. We concentrate on the predictions resulting 
from our model potentials which match observations in the \GC. For all 
of these issues, we consider the subtle effects on unequal mass binaries 
in the text (\S3); here, we focus on the results for equal mass binaries.

\begin{figure}[htb]
\centerline{\includegraphics[width=5.5in]{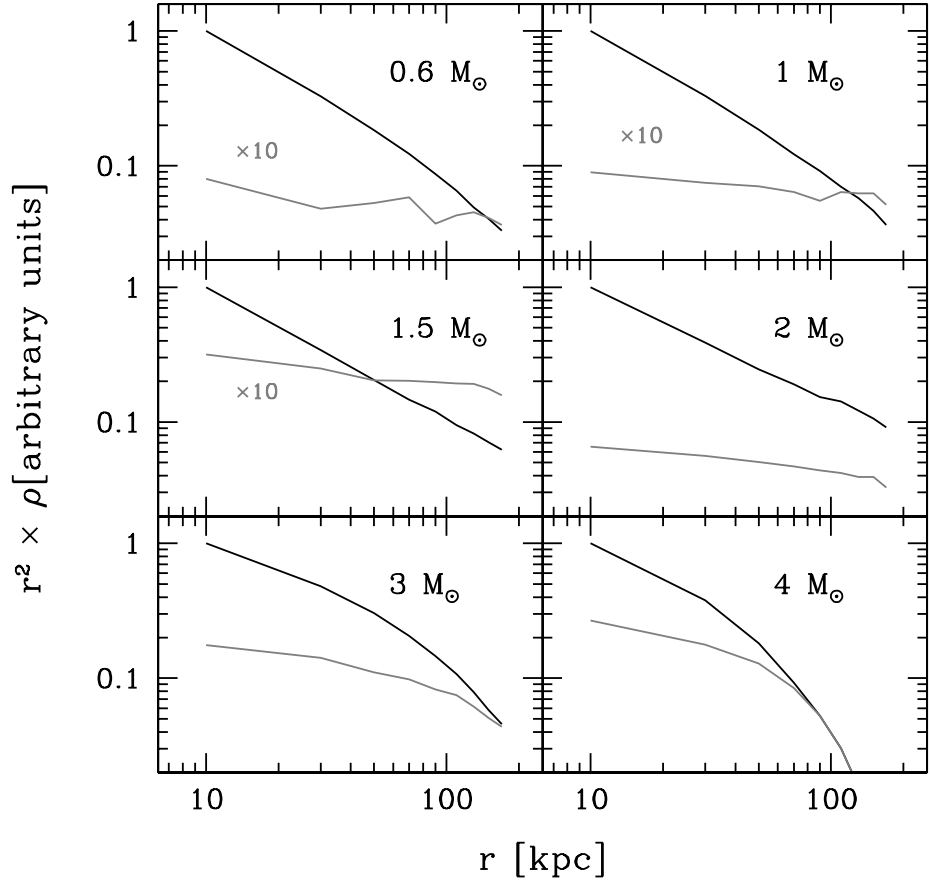}}
\caption{Relative density profiles of unbound stars ejected from equal 
mass binaries as a function of Galactocentric distance.  In each panel, 
the labels for all ejected stars (solid lines) and for unbound stars
(dashed lines) indicate the mass of the ejected stars.
\label{fig:densallhex}}
\end{figure}

\begin{figure}[htb]
\centerline{\includegraphics[width=5.5in]{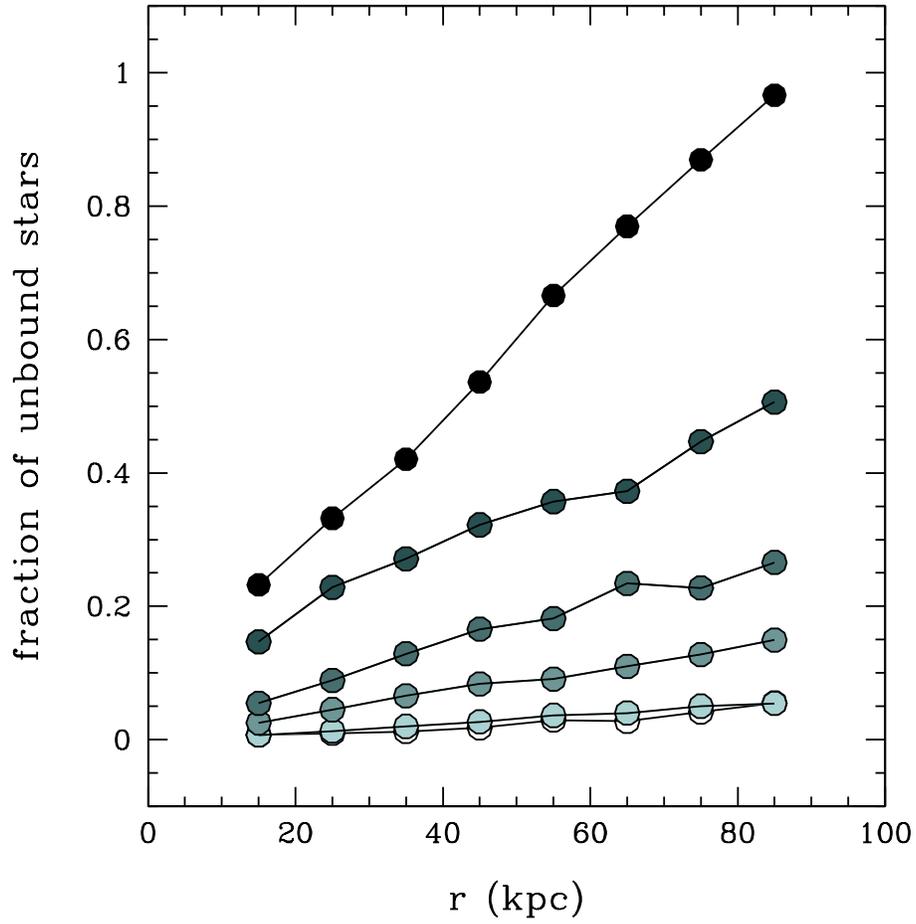}}
\caption{Fraction of unbound stars as a function of 
Galactocentric distance for equal mass binaries. The
mases are as in Fig. \ref{fig:densallhex}, with 4 \msun\ for
the top curve and 0.6 \msun\ for the bottom curve.
\label{fig:fracunbound}
}
\end{figure}

\begin{figure}[htb]
\centerline{\includegraphics[width=5.5in]{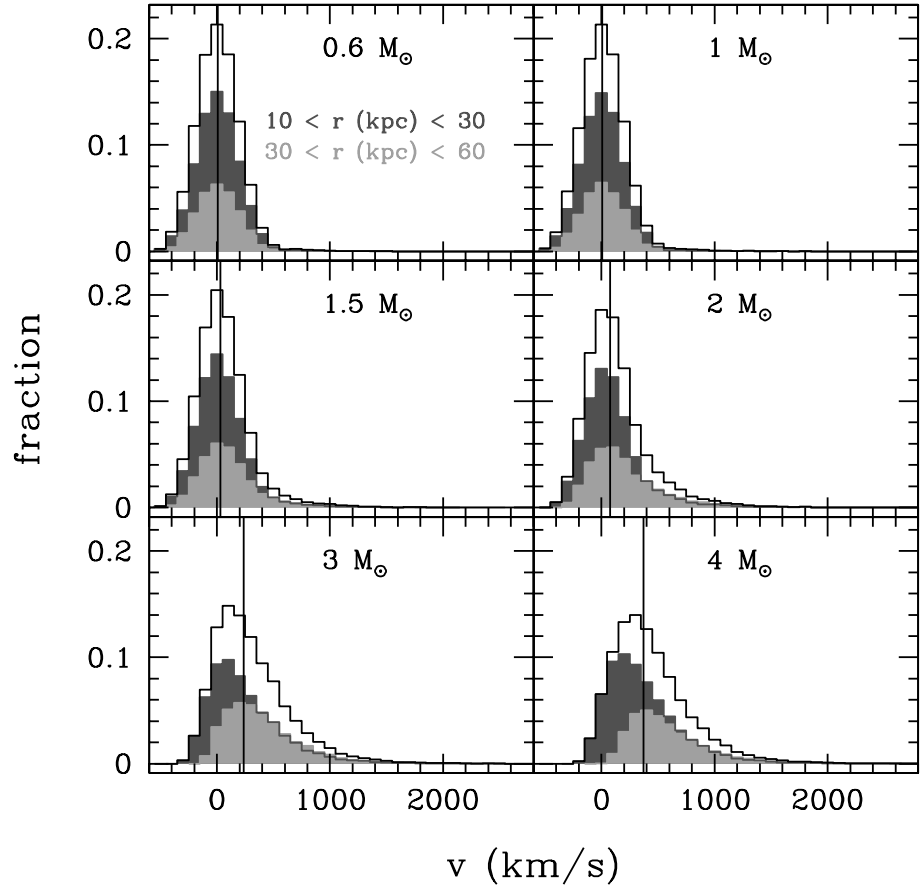}}
\caption{Predicted radial velocity distributions for equal mass binaries. 
The light grey and black histograms show the distributions in two different 
Galactocentric distance ranges; the clear histograms show results for all 
objects at 10--60~kpc.
\label{fig:fountainhex}
}
\end{figure}

\begin{figure}[htb]
\centerline{\includegraphics[width=5.5in]{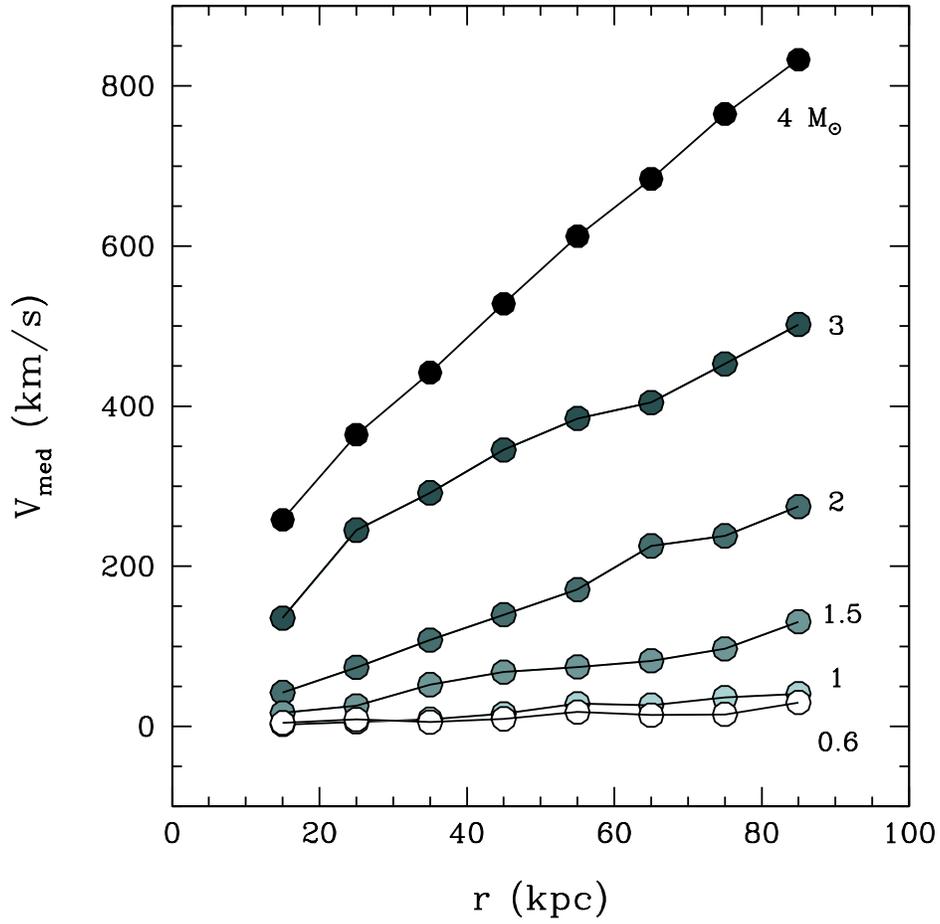}}
\caption{Median speed as a function of Galactocentric distance for stars 
ejected from equal mass binaries. The labels indicate the stellar mass; 
the shade of the points increases with mass. 
\label{fig:vvsr}
}
\end{figure}

\begin{figure}[htb]
\centerline{\includegraphics[width=5.5in]{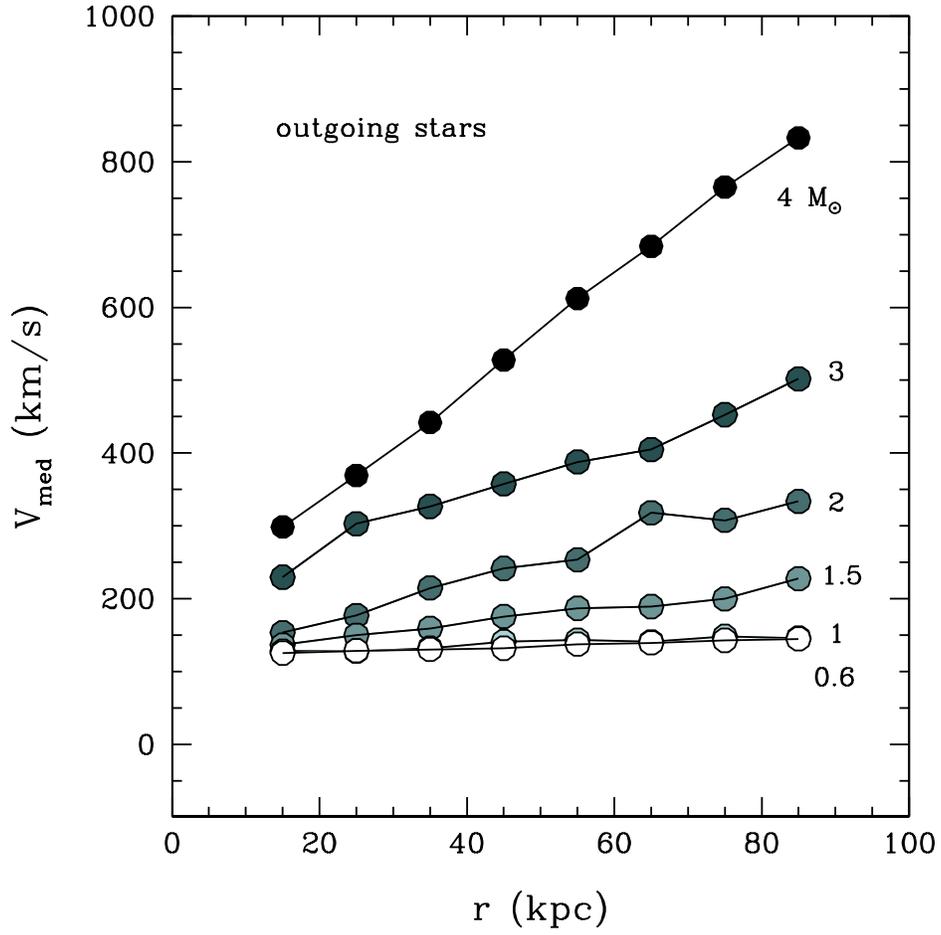}}
\caption{
As in Fig. \ref{fig:vvsr} but for outgoing stars.
\label{fig:voutvsr}
}
\end{figure}

\begin{figure}[htb]
\centerline{\includegraphics[width=5.5in]{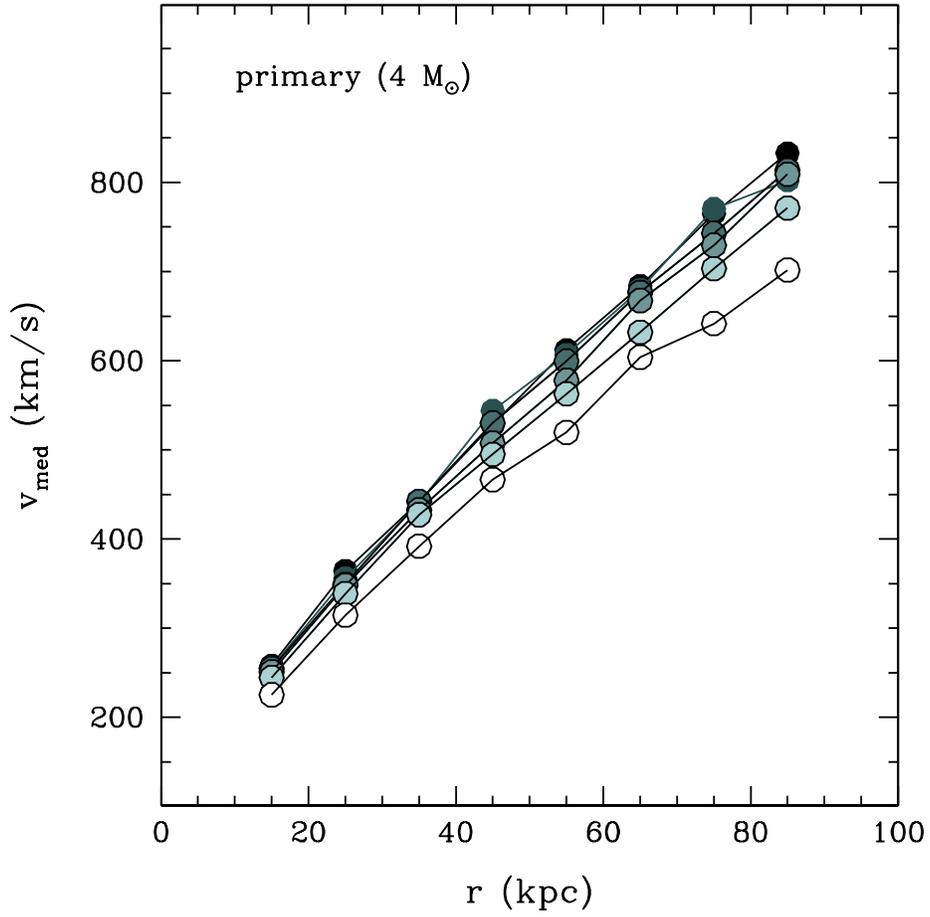}}
\caption{ 
As in Fig. \ref{fig:vvsr}, but for secondary stars in
unequal mass binaries with a 4~\Msolar\ primary.
\label{fig:uneqmvvsr1}
}
\end{figure}

\begin{figure}[htb]
\centerline{\includegraphics[width=5.5in]{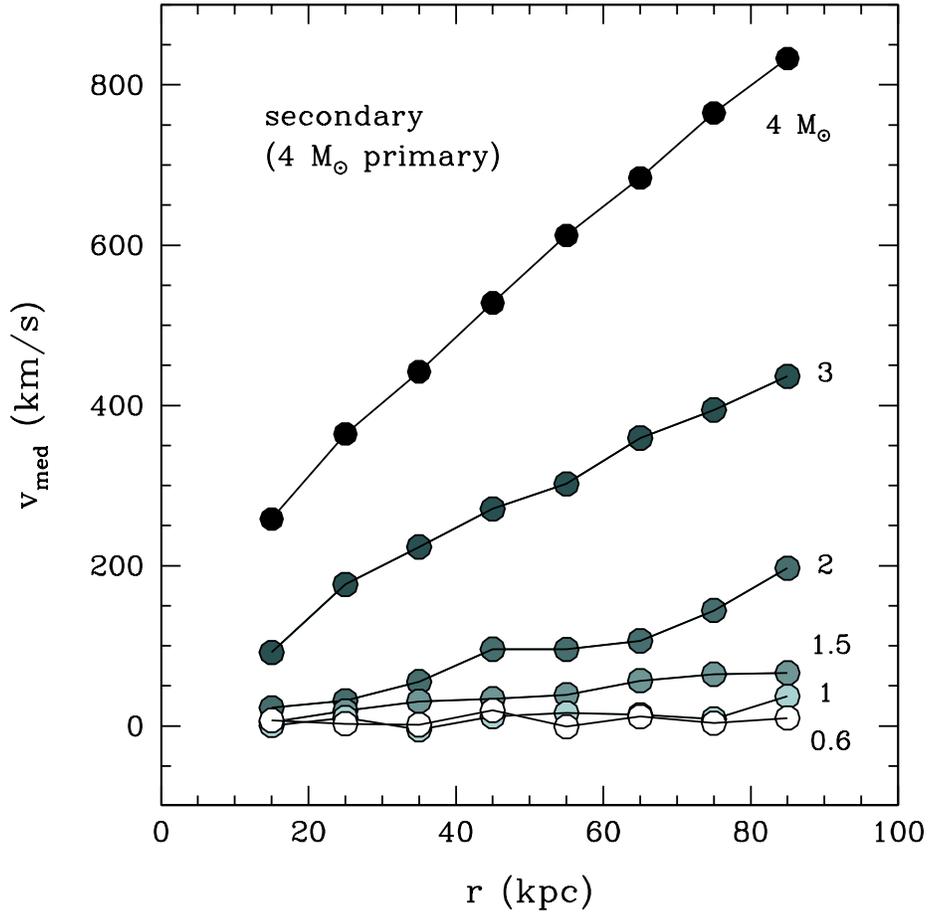}}
\caption{ 
As in Fig. \ref{fig:vvsr}, but for 4~\Msolar\ primary 
stars in unequal mass binaries.
\label{fig:uneqmvvsr2}
}
\end{figure}

\begin{figure}[htb]
\centerline{\includegraphics[width=5.5in]{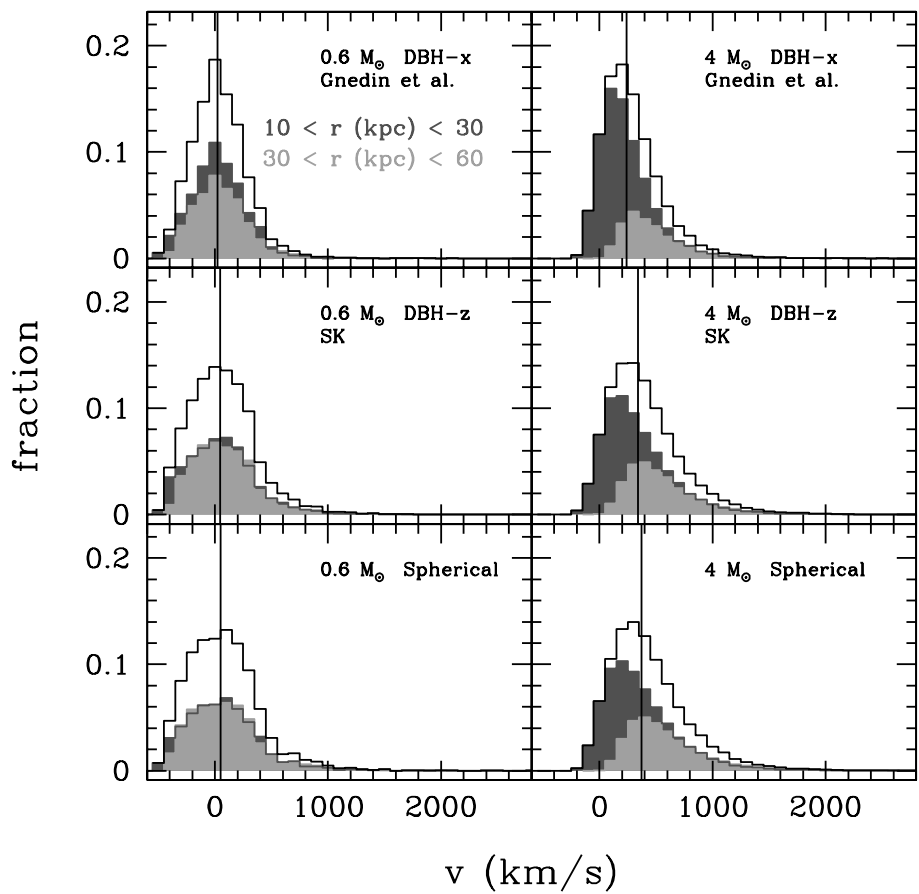}}
\caption{
As in Fig. \ref{fig:fountainhex}, but for stars in the
three component and spherical Galaxy potentials.
Upper two panels: velocity histograms in the plane of the
disk in the \citet{gne05} potential for 0.6 \msun\ stars
(left panel) and for 4 \msun\ stars. 
Middle two panels: As in the top panels for ejections
along the z-axis of our three component potential.
Lower two panels: As in the top panels for the simple
spherical potential of Figure \ref{fig:fountainhex}.
\label{fig:compmod}
}
\end{figure}

\begin{figure}[htb]
\centerline{\includegraphics[width=5.5in]{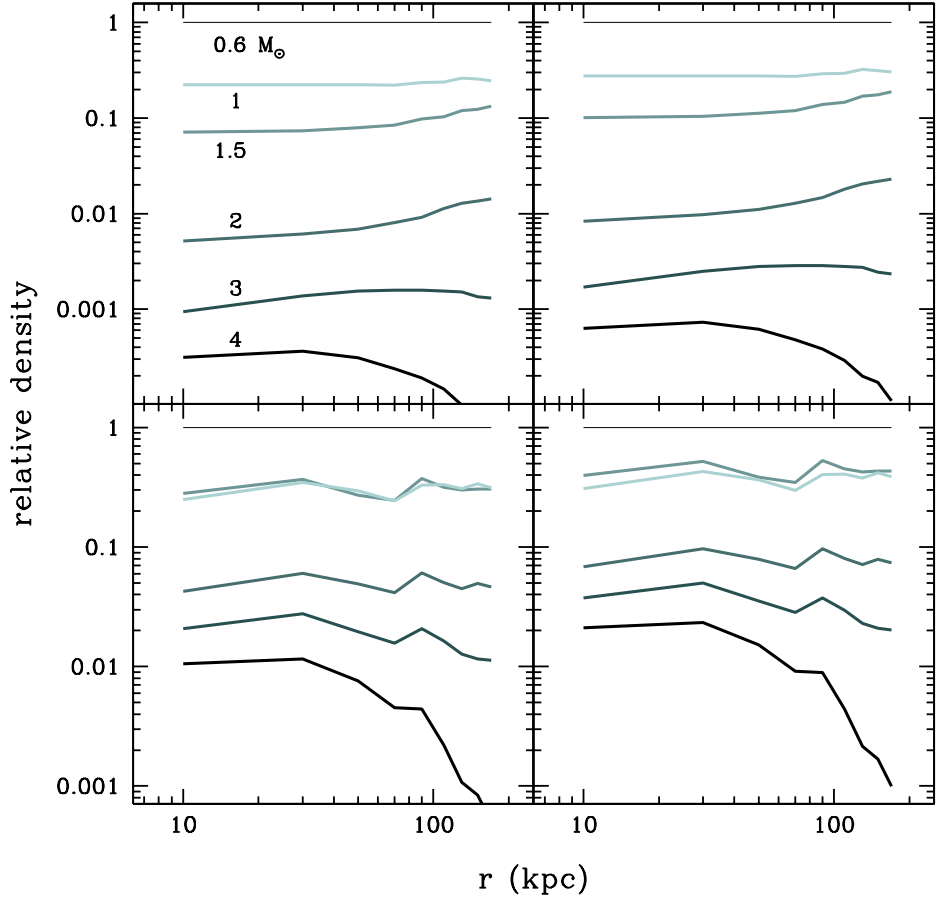}}
\caption{
Relative density profiles for HVS as a function of stellar mass. 
In all panels, we normalize the density profiles for 0.6 \msun\
stars to unity at all $r$.
Upper panels: profiles for all ejected stars for mass functions 
with $q$ = -1.35 (left panel) and $q$ = -1 (right panel).
Lower panels: as in the upper panels for unbound stars.
Compared to a population 0.6--1 \msun\ ejected stars, the relative
abundance of more massive ejected stars declines monotonically with 
stellar mass (upper panels). 
For unbound stars (lower panels), 3--4 \msun\ (1.5--3 \msun) stars 
are relatively more abundant than 0.6--1 \msun\ stars at $r \lesssim$ 
50--80 AU ($r \gtrsim$ 80--100 AU).
Although these results show that 2--4 \msun\ ejected stars are factors 
of 10--100 less numerous than 0.6--1 \msun\ ejected stars, magnitude
limited surveys for HVS are likely to yield comparable numbers of 
massive ejected stars (see Table \ref{tbl:RelNumHVS}).
\label{fig:densrelquad}
}
\end{figure}

\begin{itemize}

\item {\it Stellar evolution} affects the observability of HVSs. It removes
them at large radii where the travel time from the \GC\ exceeds the stellar 
lifetime.

Because the lifetimes of low mass stars are long, the observable properties
of low mass HVSs depend only on the potential and the properties of binaries 
at the Galactic Center. 

\item For 3--4 M$_\odot$ ejected stars, the fraction of unbound HVSs increases 
dramatically with $r$ in the Galactic halo (Fig. \ref{fig:fracunbound}). 
Because the travel time from the \GC\ is a significant fraction of the main 
sequence lifetime for these stars, the median {\it stellar age} of 
3--4 \msun\ HVSs increases monotonically with $r$.

In contrast, bound stars dominate the population of long-lived, low mass HVSs. 
Thus, the median stellar ages are essentially independent of $r$.

\item The {\it velocity distribution} of 3--4 M$_\odot$ HVSs is asymmetric
with a long tail toward positive velocities. The shape of the velocity
distribution again reflects coincidence of stellar lifetimes and travel times.

The velocity distributions of HVSs are increasingly symmetric with decreasing
stellar mass. 

\item  As emphasized by Bromley et al. (2006), the short lifetimes of 
3--4 M$_\odot$ HVS require that they be injected with large velocities 
to reach the outer halo. Thus the {\it median speed} of massive HVSs 
increases with $r$.

For low mass stars, the median speed depends only weakly on $r$. For
all stars, the median speed increases with stellar mass at fixed $r$.

\item The {\it density profiles} of unbound HVSs are approximately 
$\rho \propto r^{-n}$ with $n =$ 2--2.5.  For the most massive stars, 
their finite lifetime removes stars at large $r$, steepening the density
profile ($n \gtrsim$ 3 at $r \gtrsim$ 80 kpc).

The density profile for bound, mostly low mass HVS is roughly 
$\rho \propto r^{-3}$.  We compute the detailed behavior of 
these profiles (Fig. \ref{fig:densallhex}).

\item {\it Scaled density profiles} show that the relative numbers of 
observable unbound HVSs as a function of stellar mass are relatively 
independent of the stellar lifetime. They depend mostly on the mass 
function at the Galactic Center at the time of ejection.

In contrast, the relative numbers of bound HVSs are a function of the
stellar lifetimes and the mass function.  

\item We predict the {\it relative observable numbers} as a function of 
stellar mass for the range 0.6--4 M$_\odot$. In a magnitude limited survey,
the main factors that set detectability are (1) the accessible volume, 
(2) the fraction of unbound HVSs, (3) the asymmetry of the velocity 
distribution, and (4) contamination by indigenous stellar populations. 
Detection of HVSs is increasingly difficult with decreasing mass because
all of these issues become less and less favorable.

We also argue that post main sequence stars are poor targets because their 
detectability is subject to yet another limit, the very short lifetimes of
these phases.

\end{itemize}

Samples of HVSs which are sufficiently large to explore these predictions
provide a strong test of the model for injection of the stars into the
Galactic potential. Coupled with observations of the stellar population 
at the Galactic Center, observations of HVSs provide promising probes of 
the binary population, the stellar mass function, and the central potential 
of the Milky Way \citep[see also][]{per07b}.

So far, the observations by \citet[][and references therein]{bro07b} 
indicate that there are $\sim$ 100 detectable HVS with $m \approx$ 3--4 
M$_\odot$. Dectection of comparable lower mass populations may be 
feasible. Samples of hundreds of HVSs promise strong tests of models 
like the one we construct.

\acknowledgements

We thank J. Dubinski for helpful discussion on models for the Galactic 
potential and O. Gnedin and H. Perets for helpful comments on the 
manuscript.  We acknowledge support from the {\it NASA Astrophysics 
Theory Program} through grant NAG5-13278.

\clearpage

\begin{deluxetable}{ccccc}
\tablecolumns{5}
\tablewidth{0pc}
\tabletypesize{\footnotesize}
\tablenum{1}
\tablecaption{Adopted Properties of Main-Sequence Binary Stars}
\tablehead{
  \colhead{Mass ($M_{\odot}$)} &
  \colhead{$t_{ms}$ (Gyr)} &
  \colhead{$R_{ms}$ ($R_{\odot}$)} &
  \colhead{$a_{min,q=1}$ (AU)} &
  \colhead{$a_{min,\mb=0.6}$ (AU)} 
}
\startdata
0.6 &    13 &  0.6 & 0.015 & 0.015 \\
1   &    10 &  1.3 & 0.032 & 0.029 \\
1.5 &   2.9 &  2.5 & 0.061 & 0.051 \\
2   &   1.2 &  3.3 & 0.081 & 0.063 \\
3   &  0.35 &  4.7 & 0.115 & 0.084 \\  
4   &  0.16 &  5.5 & 0.134 & 0.100 \\
\enddata
\tablecomments{These parameters are described in \S2.1 of the text.}
\label{tbl:MSBin}
\end{deluxetable}

\begin{deluxetable}{cccccccc}
\tablecolumns{8}
\tablewidth{0pc}
\tabletypesize{\footnotesize}
\tablenum{2}
\tablecaption{Parameters for Three-Component Galaxy Potentials}
\tablehead{
  \colhead{Paper} & 
  \colhead{$M_b$ ($10^{9} M_{\odot}$)} &
  \colhead{$r_b$ (kpc)} &
  \colhead{$M_h$ ($10^{12} M_{\odot}$)} &
  \colhead{$r_h$ (kpc)} &
  \colhead{$M_d$ ($10^{10} M_{\odot}$)} &
  \colhead{$a_d$ (kpc)} &
  \colhead{$b_d$ (kpc)}
}
\startdata
DB98 &  2.50 & 0.05 & 0.8 & 21.8 & 5.0 & 2.4 & 0.18 \\
GG05 & 10.00 & 0.60 & 1.0 & 20.0 & 4.0 & 5.0 & 0.30 \\
YM07 & 10.00 & 0.50 & 1.4 & 25.9 & 4.0 & 5.0 & 0.30 \\
\\
here &  3.76 & 0.10 & 1.0 & 20.0 & 4.0 & 2.0 & 0.30 \\
\enddata
\tablecomments{These parameters are described in \S2.2 and 
Eq. \ref{eq:phis} of the text.
The cited models are DB = \citet{deh98}, GG05 = \citet{gne05}, 
YM07 = \citet{yu07}, and here = this paper}
\label{tbl:MWPot}
\end{deluxetable}

\clearpage

\begin{deluxetable}{cccccc}
\tablecolumns{6}
\tablewidth{0pc}
\tabletypesize{\footnotesize}
\tablenum{3}
\tablecaption{Mass-Radius Relation for Milky Way}
\tablehead{
  \colhead{Model} & 
  \colhead{$M_{10}$ ($10^{7} M_{\odot}$)} &
  \colhead{$M_{120}$ ($10^{9} M_{\odot}$)} &
  \colhead{$M_{210}$ ($10^{9} M_{\odot}$)} &
  \colhead{$M_{8000}$ ($10^{10} M_{\odot}$)} &
  \colhead{$M_{100000}$ ($10^{12} M_{\odot}$)}
}
\startdata
Target & 3.0 & 0.8 & 2.7 & 9.0 & 0.7 \\
\\
BKG06  & 2.9 & 1.2 & 2.2 & 9.0 & 1.1 \\
here   & 3.0 & 1.1 & 1.8 & 8.2 & 1.0 \\
\\
DB98   & 2.1 & 0.4 & 0.9 & 9.0 & 0.6 \\
WD05   & 7.0 & 1.3 & 2.7 & 9.0 & 0.7 \\
GG05   & 0.3 & 0.3 & 0.7 & 8.2 & 1.0 \\
YM07   & 0.3 & 0.3 & 0.7 & 7.8 & 1.2 \\
SGM07  & 2.3 & 1.1 & 2.4 & $\ldots$ & $\ldots$ \\
\enddata
\tablecomments{These parameters are described in \S2.2 of the text.
References are as in Table \ref{tbl:MWPot}, except for WD05 = 
\citet{wid05}, BKG06 = \citet{bkg06} and SGM07 = \citet{sto07}.}
\label{tbl:MWMass}
\end{deluxetable}

\clearpage

\begin{deluxetable}{ccccccc}
\tablecolumns{7}
\tablewidth{0pc}
\tabletypesize{\footnotesize}
\tablenum{4}
\tablecaption{Median Speeds of Ejected Stars at 30--60 kpc}
\tablehead{
  \colhead{ \ } & 
  \colhead{0.6 M$_{\odot}$} &
  \colhead{1 M$_{\odot}$} &
  \colhead{1.5 M$_{\odot}$} &
  \colhead{2 M$_{\odot}$} &
  \colhead{3 M$_{\odot}$} &
  \colhead{4 M$_{\odot}$} 
}
\startdata
\cutinhead{Simple spherical potential}
0.6 M$_\odot$&{\bf   7}&{\bf  10}&{\bf  32}&{\bf  73}&{\bf 211}&{\bf 321}\\
1   M$_\odot$&       5 &{\bf   7}&{\bf  29}&{\bf  71}&{\bf 224}&{\bf 351}\\
1.5 M$_\odot$&       5 &       3 &{\bf  32}&{\bf  77}&{\bf 229}&{\bf 361}\\
2   M$_\odot$&       5 &       5 &      19 &{\bf  77}&{\bf 235}&{\bf 364}\\
3   M$_\odot$&       1 &       6 &      15 &      49 &{\bf 238}&{\bf 369}\\
4   M$_\odot$&       6 &       3 &      18 &      39 &     179 &{\bf 372}\\
\cutinhead{Three component potential from Gnedin et al. (2005)}
0.6 M$_\odot$&{\bf   5}&{\bf   4}&{\bf  21}&{\bf  42}&{\bf 135}&{\bf 234}\\
1   M$_\odot$&       2 &{\bf   4}&{\bf  19}&{\bf  44}&{\bf 137}&{\bf 238}\\
1.5 M$_\odot$&      -1 &       1 &{\bf  17}&{\bf  42}&{\bf 137}&{\bf 240}\\
2   M$_\odot$&       3 &      -1 &       9 &{\bf  43}&{\bf 139}&{\bf 241}\\
3   M$_\odot$&       1 &       3 &      11 &      24 &{\bf 135}&{\bf 242}\\
4   M$_\odot$&       0 &       3 &       9 &      22 &     104 &{\bf 240}\\
\cutinhead{Three component potential from this paper}
0.6 M$_\odot$&{\bf   7}&{\bf  10}&{\bf  30}&{\bf  70}&{\bf 199}&{\bf 309}\\
1   M$_\odot$&       6 &{\bf   7}&{\bf  30}&{\bf  70}&{\bf 206}&{\bf 330}\\
1.5 M$_\odot$&       4 &       3 &{\bf  31}&{\bf  70}&{\bf 209}&{\bf 340}\\
2   M$_\odot$&       6 &       7 &      17 &{\bf  69}&{\bf 213}&{\bf 342}\\
3   M$_\odot$&       1 &      10 &      16 &      43 &{\bf 216}&{\bf 345}\\
4   M$_\odot$&       4 &       4 &      17 &      39 &     161 &{\bf 347}\\
\enddata

\tablecomments{Velocity data (in units of \kms) derived from simulations
with the simple spherical Galactic potential (first set of entries), the
three component Galactic potential from \citet{gne05} (second set of entries),
and the three component Galactic potential derived here (third set of entries).
The bold entries correspond to the speed of the primary star in the binary 
progenitor; otherwise the speed corresponds to the secondary star.  } 


\label{tbl:vmeds}
\end{deluxetable}
\clearpage

\begin{deluxetable}{crcccccc}
\tablecolumns{6}
\tablewidth{0pc}
\tabletypesize{\footnotesize}
\tablenum{5}
\tablecaption{Relative numbers of ejected stars}
\tablehead{
  \colhead{Mass (\msun)} & 
  \colhead{$M_V$} &
  \colhead{~~~$n_{e,21}$~~~} &
  \colhead{~~~$n_{e,23}$~~~} &
  \colhead{~~~$n_{u,21}$~~~} &
  \colhead{~~~$n_{u,23}$~~~}
}
\startdata
0.6 & 8.3~~ & 160 & 400 & 8 & 16 \\
1.0 & 5.0~~ & 125 & 300 & 6 & 20 \\
1.5 & 3.1~~ & 55 & 110 & 5 & 20 \\
2.0 & 1.7~~ & 30 & 70 & 8 & 15 \\
3.0 & $-$0.3~~ & 15 & 15 & 4 & 4 \\
4.0 & $-$0.9~~ & 1 & 1 & 1 & 1 \\
\enddata
\tablecomments{These parameters are described in \S5 of the 
main text.} 
\label{tbl:RelNumHVS}
\end{deluxetable}

\clearpage


\end{document}